\documentclass[11pt,a4paper]{article}
\pdfoutput=1
\usepackage{jheppub}
\usepackage{adjustbox}
\usepackage{slashed}
\usepackage{multirow}
\usepackage{hhline}
\usepackage{tabularx,booktabs,caption}
\usepackage{float}
\usepackage[colorlinks=false,urlcolor=cambridgeblue,linkcolor=blue,citecolor=cambridgeblue,linktocpage=true,pdfproducer=medialab,pdfa=true,anchorcolor=blue]{hyperref}
\definecolor{cambridgeblue}{rgb}{0.64, 0.76, 0.68}
\definecolor{darkraspberry}{rgb}{0.53, 0.15, 0.34}

\newcommand{\ssm}{{$\rm\Sigma SM$}}

\begin{document}

\preprint{ACFI-T20-02, LA-UR-20-22358}

\title{Collider Probes of Real Triplet Scalar Dark Matter}

\author[a,b,c]{Cheng-Wei Chiang,}
\author[d,e,a]{Giovanna Cottin,}
\author[f]{Yong Du,}
\author[g]{Kaori Fuyuto}
\author[h,f,i]{and Michael J. Ramsey-Musolf }

\affiliation[a]{Department of Physics, National Taiwan University, Taipei 10617, Taiwan}
\affiliation[b]{Institute of Physics, Academia Sinica, Taipei 11529, Taiwan}
\affiliation[c]{Physics Division, National Center for Theoretical Sciences, Taipei 10617, Taiwan}
\affiliation[d]{Departamento de Ciencias, Facultad de Artes Liberales, Universidad Adolfo Ib\'a\~{n}ez,
Diagonal Las Torres 2640, Santiago, Chile}
\affiliation[e]{Instituto de F\'isica, Pontificia Universidad Cat\'olica de Chile, Avenida Vicu\~{n}a Mackenna 4860, Santiago, Chile}
\affiliation[f]{Amherst Center for Fundamental Interactions, Department of Physics, University of Massachusetts, Amherst, MA 01003}
\affiliation[g]{Theoretical Division, Los Alamos National Laboratory, Los Alamos, NM 87545, USA}
\affiliation[h]{Tsung-Dao Lee Institute and School of Physics and Astronomy, Shanghai Jiao Tong University, 800 Dongchuan Road, Shanghai, 200240 China}
\affiliation[i]{Kellogg Radiation Laboratory, California Institute of Technology, Pasadena, CA 91125 USA}

\emailAdd{chengwei@phys.ntu.edu.tw}
\emailAdd{giovanna.cottin@uai.cl}
\emailAdd{yongdu@umass.edu}
\emailAdd{kfuyuto@lanl.gov}
\emailAdd{mjrm@sjtu.edu.cn}

%%%%%%%%%%%%%%%%%%
\abstract{We study discovery prospects for a real triplet extension of the Standard Model scalar sector at the Large Hadron Collider (LHC) and a possible future 100\,TeV $pp$ collider. We focus on the scenario in which the neutral triplet scalar is stable and contributes to the dark matter relic density. When produced in $pp$ collisions, the charged triplet scalar decays to the neutral component plus a soft pion or soft lepton pair, yielding a disappearing charged track in the detector.
We recast current 13\,TeV LHC searches for disappearing tracks, and find that the LHC presently excludes a real triplet scalar lighter than 248 (275) GeV, for a mass splitting of 172 (160) MeV with $\mathcal{L}=\rm36\,$fb$^{-1}$.
{\textcolor{black}{The reach can extend to 497 (520)\,GeV with the collection of  $3000\,$fb$^{-1}$.}} We extrapolate the 13\,TeV analysis to a prospective 100\,TeV $pp$ collider, and find that a $\sim3$\,TeV triplet scalar could be discoverable with $\mathcal{L}=30$ ab$^{-1}$, depending on the degree to which pile up effects are under control. We also investigate the dark matter candidate in our model and corresponding present and prospective constraints from dark matter direct detection. We find that currently XENON1T can exclude a real triplet dark matter lighter than $\sim3$\,TeV for a Higgs portal coupling of order one or larger, and the future XENON20T will cover almost the entire dark matter viable parameter space except for vanishingly small portal coupling.}

% 287\,GeV with $\mathcal{L}=\rm36\,$fb$^{-1}$.  The reach will extend to 608\,GeV and 761\,GeV with the collection of  $\mathcal{L}=300\,$fb$^{-1}$  and $3000\,$fb$^{-1}$ respectively. 

\arxivnumber{2003.07867}

\maketitle
\flushbottom

%%%%%%%%%%%%%%%%%%%%%%
\section{Introduction} \label{intro}
%%%%%%%%%%%%%%%%%%%%%%
Deciphering the identity of dark matter (DM) is one of the primary ambitions in particle physics. The existence was first hypothesized to account for the motion of galaxies in clusters\,\cite{Zwicky:1933gu}, and subsequently established by various cosmological observations (for a recent review, see\,\cite{Bertone:2004pz, Battaglieri:2017aum}). The latest measurements of the cosmic microwave background anisotropies show that the energy density of the dark matter is $\Omega_{\rm DM} h^2=0.1198\pm0.0012$\,\cite{Aghanim:2018eyx} with the Hubble parameter $h$ in units of $100~ {\rm km}/{\rm (s\cdot Mpc)}$. None of the Standard Model (SM) particles can satisfy the DM properties, pointing to new physics beyond it. So far, a plethora of theoretical models have been proposed, indicating a wide mass range of DM candidates from $10^{-15}~$GeV to $10^{15}$~GeV. Nevertheless, in recent years, a diverse range of experimental ideas have been proposed while the existing DM searches have significantly upgraded their experimental sensitivities. 

Weakly interacting massive particles (WIMPs), whose mass range is roughly between $10~$GeV and a few\,TeV, have long been considered an appealing DM candidate. The WIMP scenario assumes that DM particles are initially in thermal and chemical equilibrium, and then, freeze out at some point as the Universe expands. A widely discussed realization of the WIMP scenario is supersymmetry, where the lightest neutralino becomes a DM candidate. Another viable WIMP candidate is a neutral component of an electroweak multiplet (singlet under SU(3)$_C$). A comprehensive study of all possible electroweak multiplets has been done in\,\cite{Cirelli:2005uq}, and the related phenomenology has been studied in a multitude of works\,\cite{Pal:1987mr, FileviezPerez:2008bj, Hambye:2009pw, Fischer:2013hwa, JosseMichaux:2012wj, Basak:2012bd, Araki:2011hm, AbdusSalam:2013eya, Lu:2016dbc, Chao:2018xwz, Abe:2014gua, Cirelli:2014dsa, Harigaya:2015yaa, Matsumoto:2017vfu, Cirelli:2009uv, Cao:2018nbr, DiLuzio:2018jwd, Chua:2013zpa, Kadota:2018lrt, Cirelli:2015bda, Cai:2017wdu, Cai:2015kpa, Xiang:2017yfs, Chigusa:2018vxz, Matsumoto:2018ioi, Kuramoto:2019yvj, Abe:2019egv}. Among these scenarios is a real $SU(2)_L$ triplet scalar $(\Sigma)$ with a zero hypercharge $(Y=0)$, which is the simplest extension of the SM scalar sector involving particles carrying electroweak charge. In this model -- the \ssm -- imposing a $Z_2$ symmetry enables the neutral component $(\Sigma^0)$ to be stable. Previous works have shown that the correct thermal relic abundance is obtained if the mass of the neutral component is around $2.5~$TeV. For this mass regime,  the corresponding search at the Large Hadron Collider (LHC) is challenging.

Nevertheless, the previous study\,\cite{FileviezPerez:2008bj} discussed the possibility of distinctive charged track events at the LHC. In the \ssm, the $(\Sigma^{\pm})$ and neutral scalars are degenerate at tree level. However, a one-loop radiative correction generates a small mass splitting {\color{black}$\Delta m\simeq166~$MeV\,\cite{Cirelli:2005uq}, which gets further modified by a few MeV if two-loop corrections are also included\,\cite{Ibe:2012sx}}. In this case, the charged scalar becomes a relatively long-lived particle. If such a long-lived charged particle has a decay length of $O(1)$ cm, it can leave a disappearing track in detectors. The main decay mode of the charged triplet scalar is $\Sigma^{\pm}\to\Sigma^0\pi^{\pm}$, which results in a decay length $c\tau_{\Sigma^{\pm}}=5.06~$cm\,\cite{Cirelli:2005uq}. Thus, the disappearing track searches have the great potential to observe the signature of the charged particle. The same strategy has comprehensively been discussed to search for compressed dark sectors \cite{Mahbubani:2017gjh}, neutralino DM at the LHC\,\cite{Aaboud:2017mpt} and future hadron collider\,\cite{Han:2018wus, Saito:2019rtg, Fukuda:2017jmk, Mahbubani:2017gjh}.   

In this work, we explore the discovery reach for the triplet scalar DM with a disappearing charged track (DCT) signature at the LHC and a prospective future $100$\,TeV $pp$ collider. We pay particular attention to the triplet interaction with the SM Higgs doublet. Previous studies \cite{Cirelli:2005uq,Cirelli:2007xd} have neglected the corresponding Higgs portal coupling, whose presence may modify both the DM and collider analyses in the following ways: $1)$ annihilation cross sections of the DM, $2)$ the DM-nucleon spin-independent elastic cross section, and $3)$ production cross sections of the charged scalars. Our analysis not only updates the possibility of the DM candidate in the \ssm\, taking into account the nonzero Higgs portal coupling, but also investigates the reach of a DCT search at the 13\,TeV LHC and provides a rough estimate at a future 100\,TeV hadron collider.  In undertaking our LHC DCT analysis, we first validate our approach by recasting the ATLAS search for disappearing tracks in Ref.\,\cite{Aaboud:2017mpt}. In making projections for a prospective 100 TeV collider, we also take into account present uncertainty about the impact of pileup effects, drawing on the work of Ref.~\cite{Saito:2019rtg}. Our treatment of the DM dynamics entails solving the relevant Boltzmann equations, including effects of coannihilation and Sommerfeld enhancement. We find that
\begin{itemize}
\item Utilizing the DCT signature, the LHC with $\sqrt{s}=13$\,TeV {\color{black}and $\mathcal{L}=36\rm\,fb^{-1}$ excludes a real triplet lighter than $\sim248\,(275)$\,GeV for $\Delta m=172\,(160)$\,MeV}, under the assumption of a $Z_2$-symmetry in the corresponding scalar potential. {\color{black}For $\mathcal{L}=300\rm\,fb^{-1}$ and $\mathcal{L}=3000\rm\,fb^{-1}$, we find the prospective future LHC exclusion reach is $\sim535\,(590)$\,GeV and $\sim666\,(745)$\,GeV optimistically, and $\sim348\,(382)$\,GeV and $\sim496\,(520)$\,GeV with the inclusion of a 30\% systematic uncertainty for $\Delta m=172\,(160)$\,MeV}. A future 100\,TeV $pp$ collider could discover a real triplet up to $\sim3$\,TeV with $\mathcal{L}=30\rm\,ab^{-1}$. {However, the precise reach of a 100\, TeV collider depends significantly on assumptions about pileup effects. Discovery of the \ssm\, over the entire region of DM-viable parameter space would require that such pileup effects are under sufficient control.} We show our results in figure\,\ref{fig:13TeVlimit} and table\,\ref{table:Nsignal}.
\item {For a Higgs portal coupling of $\mathcal{O}(2.5)$ or larger}, XENON1T rules out  real triplet DM lighter than $\sim3$\,TeV. The future XENON20T will be able to explore almost the entire DM parameter space except for a {vanishingly small} Higgs portal coupling. We present our results in figure\,\ref{dmdirdet}.
\end{itemize}

This paper is organized as follows. We discuss the basic structure of the \ssm\, in secton\,\ref{setup} and show our analysis on the disappearing track at the LHC and a 100\,TeV $pp$ collider in secton\,\ref{collider}. We then present the DM relic density and DM direct detection constraints in secton\,\ref{tripletdm}. Finally, we summarize our conclusions in secton\,\ref{sec:summary}.

%%%%%%%%%%%%%%%%%%%%%%
\section{The Real Triplet Model (\ssm)}\label{setup}
%%%%%%%%%%%%%%%%%%%%%
\subsection{\ssm\, setup}
The scalar sector Lagrangian for the \ssm\, is given by
\begin{align}
{\cal L}=\left(D_{\mu}H \right)^{\dagger}\left(D^{\mu} H\right)+\left(D_{\mu}\Sigma \right)^{\dagger}\left(D^{\mu}\Sigma\right)-V\left(H,\Sigma \right),
\end{align}
where the $SU(2)$ doublet Higgs $H$ and triplet scalar $\Sigma$ are cast into the form
\begin{align}
H=
\begin{pmatrix}
G^+\\
\frac{1}{\sqrt{2}}\left(v+h+iG^0\right)
\end{pmatrix},\hspace{1cm}
\Sigma=\frac{1}{2}
\begin{pmatrix}
\Sigma^0 & \sqrt{2}\Sigma^+\\
\sqrt{2}\Sigma^- & -\Sigma^0
\end{pmatrix},
\end{align}
with the Higgs vacuum expectation value (VEV) $v\simeq 246~$GeV.
The covariant derivative acting on $\Sigma$ is defined by $D_{\mu}\Sigma=\partial_{\mu}\Sigma+ig_2\left[W_{\mu},\Sigma \right]$ with the product of the $SU(2)$ gauge boson and Pauli matrices $W_{\mu}=W^a_{\mu}\tau^a/2$ (the corresponding expression for $D_\mu H$ is standard). The scalar potential is expressed by
\begin{align}
V\left(H, \Sigma \right)=-\mu^2H^{\dagger}H+\lambda_0\left(H^{\dagger}H \right)^2-\frac{1}{2}\mu^2_{\Sigma} F+\frac{b_4}{4}F^2+\frac{a_2}{2}H^{\dagger}HF,
\end{align}
where $F=\left(\Sigma^0 \right)^2+2\Sigma^+\Sigma^-$. In the above potential, we impose a $Z_2$ discrete symmetry in which $\Sigma$ transforms with a $Z_2$-odd parity while all the others are $Z_2$-even. Therefore, the scalar trilinear term $H^{\dagger}\Sigma H$ is forbidden. The scalar masses are given by
\begin{align}
m^2_{h}=2\lambda_0v^2,\quad m^2_{\Sigma^{0}}=m^2_{\Sigma^{\pm}}=-\mu^2_{\Sigma}+\frac{a_{2}v^2}{2}\equiv m^2_0.
\end{align}
Although the charged and neutral components of $\Sigma$ are degenerate at tree level, the degeneracy is broken by an electroweak radiative correction to the mass terms. Depending on $m_0$, the mass difference is given by\,\cite{Cirelli:2005uq}
\begin{align}
\Delta m=m_{\Sigma^{\pm}}-m_{\Sigma^0}=\frac{\alpha_2m_0}{4\pi}\left[f\left(\frac{m_W}{m_0} \right)-c^2_Wf\left(\frac{m_Z}{m_0} \right) \right],
\end{align}
where $\alpha_2=g^2_2/\left(4\pi\right)$, $m_{W(Z)}$ is the $W(Z)$ boson mass, $c_W=\cos\theta_W$ is the cosine of the weak mixing angle, and $k$ is a UV regulator. The loop functions are
\begin{align}
f\left(r \right)=-\frac{r}{4} \left[2r^3\log r-kr+\left(r^2-4 \right)^{\frac{3}{2}}\ln A\left(r\right) \right],
\end{align}
where
\begin{align}
A\left(r\right)=\frac{1}{2}\left(r^2-2-r\sqrt{r^2-2} \right).
\end{align}
In the case of $m_0\gg m_W$, the above expression can be simplified, leading to $\Delta m=\left(166\pm1 \right)$\,MeV. This mass splitting ensures the decay channel  $\Sigma^{\pm}\to \Sigma^{0}\pi^{\pm}$ is kinematically allowed, and the corresponding rate is
\begin{align}
\Gamma\left(\Sigma^{\pm}\to\Sigma^{0}\pi^0 \right)=\frac{2G_F^2}{\pi}f^2_{\pi}V_{ud}^2\left(\Delta m \right)^3\sqrt{1-\frac{m^2_{\pi}}{\left(\Delta m\right)^2}}\label{sig:decayrate},
\end{align}
where the other quantities in this expression are the Fermi constant $G_F$, pion decay constant $f_{\pi}~(=131~{\rm MeV})$, the CKM matrix $V_{ud}$ and pion mass $m_{\pi}$.\footnote{The expression corresponds to the leading term of an expansion with respect to $\Delta m/m_{\Sigma^{\pm}}$, in which dependence on $m_{\Sigma^{\pm}}$ is canceled out and only the mass difference remains.} This decay mode accounts for $98\%$ of the branching ratio, and the remaining modes are $\Sigma^{\pm}\to \Sigma^{0}\mu^{\pm}\nu_{\mu}$ and $\Sigma^{\pm}\to \Sigma^{0}e^{\pm}\nu_e$. And, it follows that the charged scalar has a relatively long lifetime $\tau_{\Sigma^{\pm}}\sim 0.17~$ns.

%%%%%%%%%%%%%%%%%%%%%%
\subsection{Phenomenological aspects}
%%%%%%%%%%%%%%%%%%%%%%
\begin{figure}[t]
\begin{center}
\includegraphics[width=10cm]{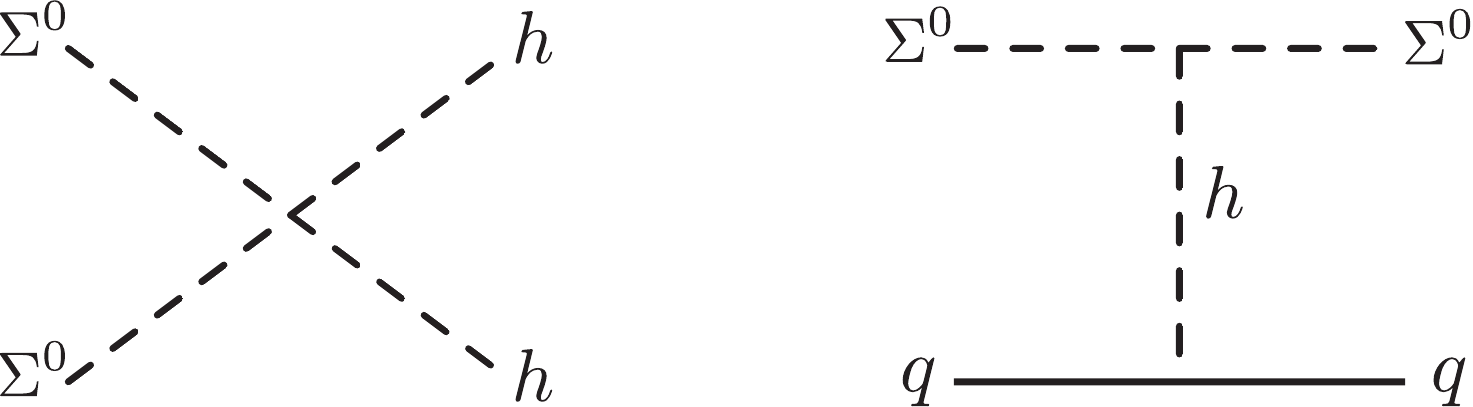} 
\end{center}
\caption{Examples of contributions from the Higgs portal coupling to the DM annihilation (left) and spin-independent (right) cross section. The variable $q$ represents the SM quarks.}
\label{fig:DMportal}
\end{figure}

\begin{figure}[t]
\begin{center}
\includegraphics[width=12cm]{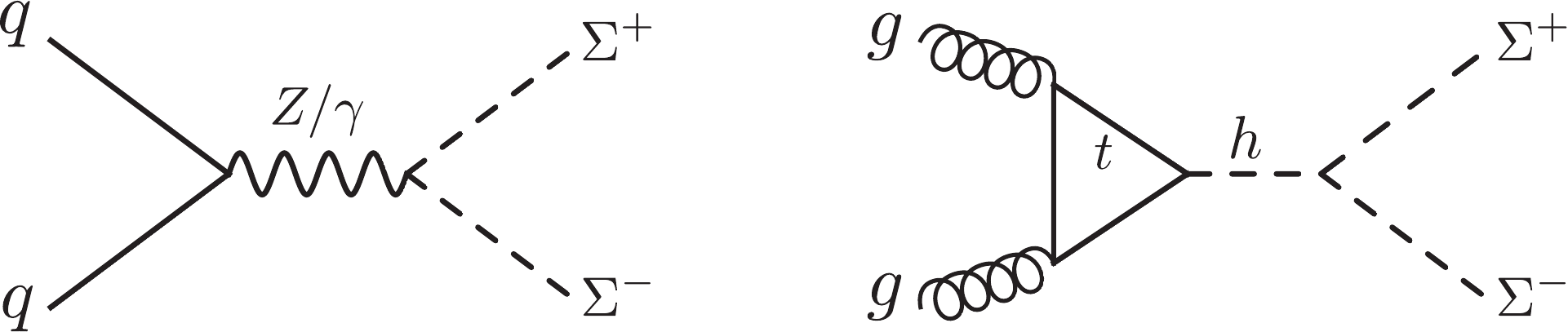} 
\end{center}
\caption{Examples of production mechanisms of the charged scalars : the DY (left) and $ggF$ (right) processes.}
\label{fig:Colliderportal}
\end{figure}

Here, we briefly remark on two main points of our study:

\begin{itemize}
\item{DM candidate : $\Sigma^0$}

In this model, the neutral scalar $\Sigma^0$ can be a DM candidate. In order to render the neutral scalar stable, in addition to the $Z_2$ symmetry, a triplet VEV $\langle \Sigma\rangle$ should not develop. Otherwise, the Higgs portal interaction in the scalar potential yields mixing with the SM Higgs and allows $\Sigma^0$ to decay into the SM particles. Assuming that $\Sigma^0$ {saturates} the observed DM abundance,  previous studies showed that the mass of this DM candidate must be around $2.5$\,TeV\,\cite{Cirelli:2005uq,Cirelli:2007xd}. However, the results have been obtained by neglecting the Higgs portal coupling $a_2$. Once the portal coupling becomes nonzero, it yields new contributions to annihilation cross section of the DM and the DM-nucleon spin-independent cross section as in Fig. \ref{fig:DMportal}. In a recent study of electroweak multiplet dark matter for higher dimensional representations of SU(2)$_L$, it was shown that inclusion of the non-vanishing Higgs portal coupling can substantially alter the relationship between the relic density and dark matter mass\cite{Chao:2018xwz}. Consequently, in the following section, we update the analyses of the relic density including the dependence on $a_2$.

\item{Disappearing track search : $\Sigma^{\pm}\to\Sigma^0\pi^{\pm}$}

The small mass splitting between $\Sigma^{\pm}$ and $\Sigma^0$ gives a smoking gun signature of a DCT, which has been searched for at the LHC \cite{Khachatryan:2016sfv,Aaboud:2017mpt}. The charged scalar can travel a macroscopic distance before decaying into the neutral scalar and a pion, which may leave multiple hits in the tracking layers. The produced pion has a very low momentum $(\sim 100{\rm\,MeV})$; therefore it is too soft to be reconstructed, leading to a signature of a track that disappears. The previous study in \cite{FileviezPerez:2008bj} analyzed disappearing track events in the electroweak Drell-Yann (DY) process (left diagram of Fig. \ref{fig:Colliderportal}) with a single initial state radiation. The authors concluded that one could expect to see several hundred track events in $100~$fb$^{-1}$ at the LHC. However, for the DM mass range considered in that work the $\Sigma^0$ can explain only a portion of the present relic density. In the presence of the Higgs portal coupling, an additional production mechanism, gluon-gluon fusion process $(ggF)$ in the right diagram of Fig. \ref{fig:Colliderportal}, can increase the number of disappearing track events. In what follows, including the $ggF$ process, we analyze the reach with disappearing track searches, including a mass range for $m_{\Sigma^0}$ consistent with the observed relic density.
\end{itemize}

%%%%%%%%%%%%%%%%%%%%%%
\section{Collider phenomenology with disappearing track searches}\label{collider}
%%%%%%%%%%%%%%%%%%%%%%

ATLAS can currently reconstruct tracks as short as $\mathcal{O}(10)$ cm, providing the opportunity to search for long-lived particle with lifetimes of $\mathcal{O}(0.2)$ ns\cite{Aaboud:2017mpt}.\footnote{For a state-of-art review on long-lived particle searches at the LHC, see Ref.\,\cite{Alimena:2019zri}}. We, therefore, study the discovery potential of the \ssm\, at the LHC by recasting the ATLAS search for disappearing tracks reported in Ref.\,\cite{Aaboud:2017mpt}. {\color{black}We also provide optimistic projections for the High-Luminosity LHC (HL-LHC) and a rough extrapolation of the reach to a hypothetical 100\,TeV collider. We adopt a benchmark set of parameters yielding $c\tau=68.42, 55.36, 46.11$ mm throughout our study, consistent with $\Delta m = 160, 166, 172$\,MeV, respectively\footnote{{\color{black}Our choice of mass splittings is motivated through considering two-loop corrections to the mass splitting. See, for instance, Ref.\,\cite{Ibe:2012sx}}.}.}

\begin{figure}[t]
\centering{
  \begin{adjustbox}{max width = \textwidth}
\begin{tabular}{cc}
\includegraphics[scale=0.35]{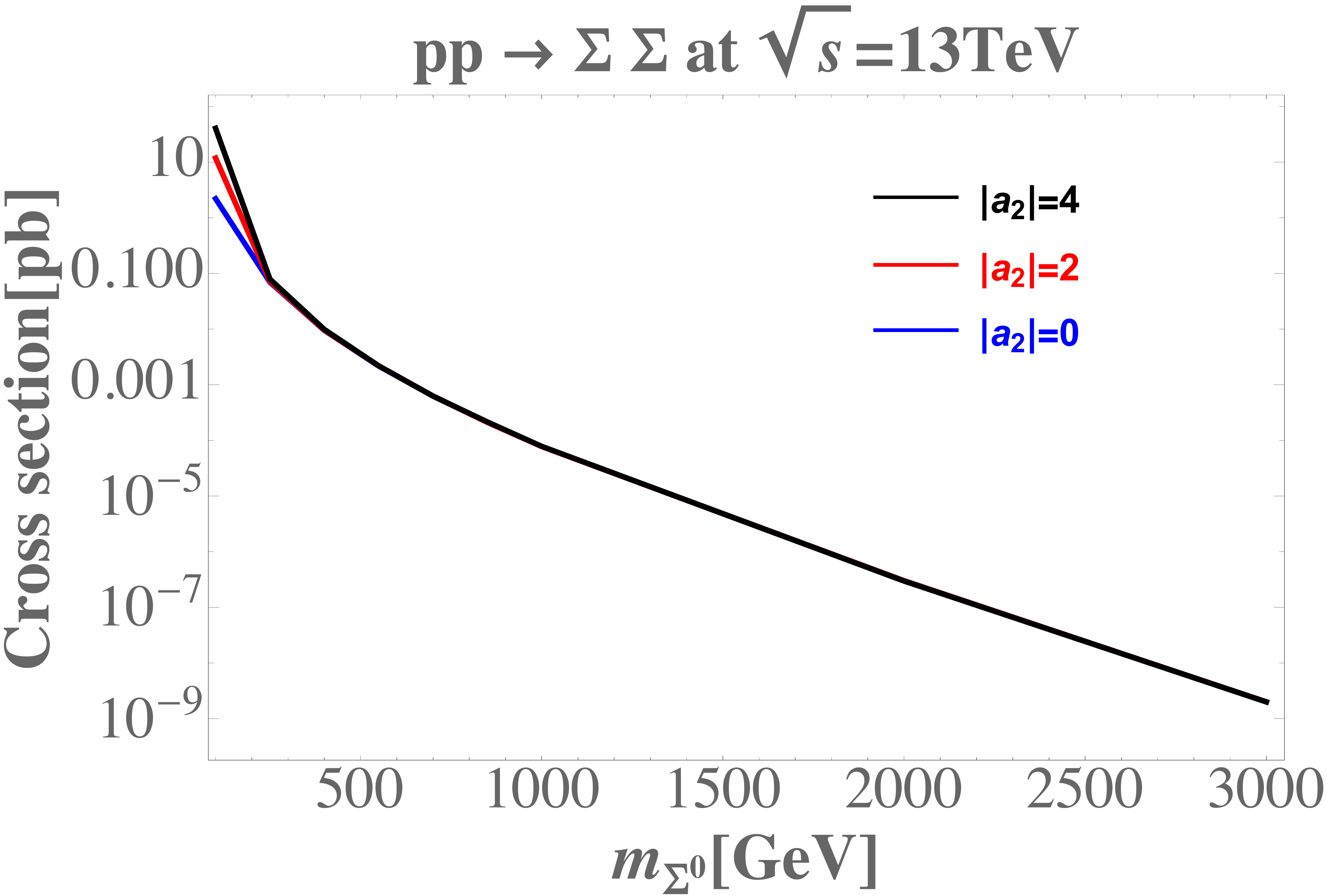} ~& ~ \includegraphics[scale=0.35]{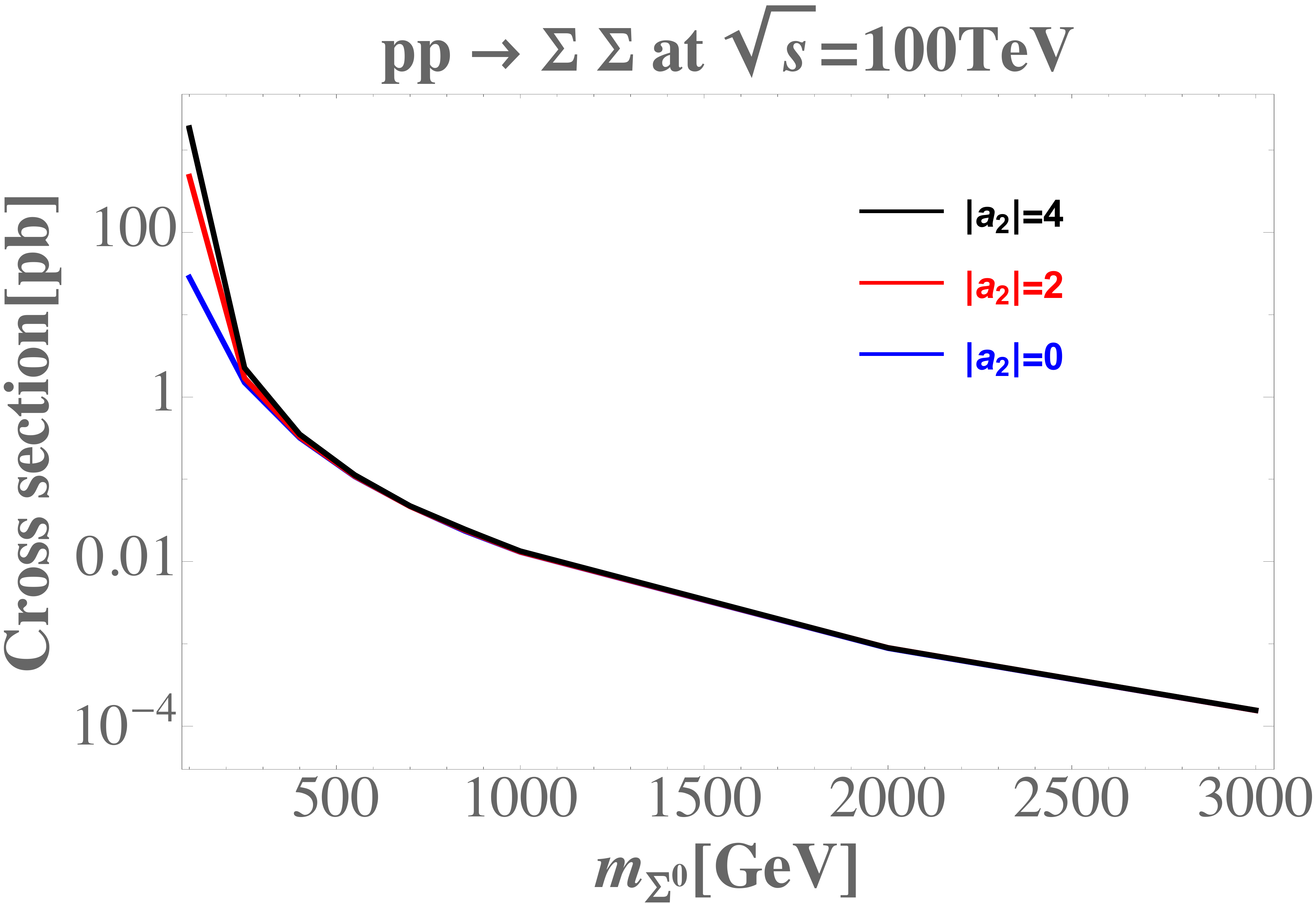}
\end{tabular}
  \end{adjustbox}}
\caption{Pair production cross sections of triplet particles $\Sigma=\Sigma^{\pm,0}$ at 13\,TeV and 100\,TeV $pp$ colliders as a function of $m_{\Sigma^0}$ with representative values of $a_2$.}
\label{fig:sigma}
\end{figure}

figure\,\ref{fig:sigma} shows the pair production cross-sections for $pp\rightarrow \Sigma\Sigma$, with $\Sigma=\Sigma^{\pm,0}$, at both 13\,TeV and 100\,TeV colliders calculated with {\tt MadGraph2.6.1}\,\cite{Alwall:2014hca}.\footnote{We compute cross sections at the LO at both 13\,TeV and 100\,TeV. The NLO effects are very modest, with a $K-$factor of 1.18 at the 13\,TeV LHC. See, e.g. the discussion in Ref.\,\cite{Ramsey-Musolf:2019lsf,Fuks:2013lya,Fiaschi:2019zgh}.} Note that the cross sections have some $a_2$ dependence only when $m_{\Sigma^0}\lesssim300\rm\,GeV$ and that $m_{\Sigma}\lesssim90$\,GeV has already been excluded by LEP\,\cite{Egana-Ugrinovic:2018roi}.\footnote{LEP places a combined lower limit on chargino mass, for example, at 92.4\,GeV\,\cite{lepsusy}.} The $a_2$ dependence in figure\,\ref{fig:sigma} can be understood as follows:
\begin{itemize}
\item For the $\Sigma^\pm\Sigma^0$ final state, it is uniquely produced through $qq'\to W^{\pm*}\to\Sigma^\pm\Sigma^0$, and there is no $a_2$ dependence.
\item For the $\Sigma^+\Sigma^-$ ($\Sigma^0\Sigma^0$) final state, the production channels are $gg/q\bar{q}\to h^{*}/\gamma^*,Z^* \to\Sigma^+\Sigma^-$ ($gg/q\bar{q}\to h^{*}\to\Sigma^0\Sigma^0$), where the $a_2$ dependence arises from the $h\Sigma^+\Sigma^-$ ($h\Sigma^0\Sigma^0$) vertex. However, when the triplet becomes heavy such that the square of the parton center of mass energy $\hat{s}>4m_t^2$, where $m_t$ is the top quark mass, the $ggh$ form factor decreases dramatically such that the Drell-Yan processes dominate. In this regime, we thus lose the $a_2$ dependence.
\end{itemize}

We point out that the pair production cross sections given in figure\,\ref{fig:sigma} are all calculated at the leading order (LO) with {\tt MadGraph2.6.1}. {As discussed in Ref.\,\cite{Dolan:2012ac}, next-to-leading-order (NLO) QCD corrections could enhance the cross section by a factor of about 2 for the ggF process. Therefore, our production cross section above for the $\Sigma^{\pm(0)}\Sigma^{\mp(0)}$ processes {for $|a_2|\sim\mathcal{O}(1)$ or larger is an underestimate when the triplet is light. When the triplet is heavy, which is relevant for our DM study as detailed below, since the ggF process will be suppressed as discussed above, the most relevant NLO QCD corrections are those applicable to the electroweak Drell-Yan process. As summarized in Ref.\,\cite{Ramsey-Musolf:2019lsf}, the corresponding K-factor is about 1.18 for the LHC with $\sqrt{s}=13$\,TeV, which corresponds to mild corrections to our LO results. Thus, we do not expect the NLO QCD corrections to have a substantial impact on our analysis of the LHC sensitivity. On the other hand, since the corresponding K-factor for a future 100\,TeV collider does not exist, we will not include the corresponding corrections in our analysis of the higher energy $pp$ Drell-Yan process.}
}

In what follows, we present the recast details of the ATLAS search for disappearing tracks in Ref.\,\cite{Aaboud:2017mpt}.

%%%%%%%%%
\subsection{Validation of the ATLAS 13\,TeV disappearing track search}\label{atlasval}
%%%%%%%%%
The ATLAS 13\,TeV search in Ref.\,\cite{Aaboud:2017mpt} looks for long-lived charginos based 
on a DCT signature. To make sure the calibration of our simulations is reliable before its application to the \ssm, we first validate the ATLAS result for their electroweak anomaly-mediated supersymmetry breaking (AMSB) benchmark model. Events are generated with {\tt MadGraph2.6.1}\,\cite{Alwall:2014hca} and showered with
{\tt Pythia8}\,\cite{Sjostrand:2014zea}. Our detector simulation is based on a custom made code which replicates the ATLAS 13\,TeV search. 

The ATLAS search selects events with large missing transverse momentum ($\slashed{p}_T$), and the signal topology targeted is characterized to have a high-$p_{T}$ jet to ensure large {$\slashed{p}_T$}. A candidate event is required to have at least one ``pixel tracklet'', which is a short track with only pixel hits (i.e with no associated hits in the strip semiconductor tracker or SCT). Furthermore, the candidate pixel tracklets are required to have $p_{T}>100$\,GeV. In Ref.\,\cite{Aaboud:2017mpt}, the authors interpreted the result in the context of  AMSB for both electroweak and strong production of charginos. We use the efficiency maps directly on Monte Carlo truth information (i.e., generator-level chargino decay position, $\eta$ and $p_{T}$), as we can not simulate the tracklet's quality requirements and disappearance condition.

Backgrounds for disappearing tracks can arise from charged particles scattered by
the material and fake tracks. The ATLAS search in\,\cite{Aaboud:2017mpt} provides a functional form for the $p_{T}$ distribution of fake tracklets, which can be used to estimate the fake tracklet background. We do not perform any background estimation in this article given the complexity of the estimation. Instead, we compare with the ATLAS model independent upper limit on the cross section in secton\,\ref{LHC13TeV} for the 13\,TeV case.  For the 100\,TeV case, we use the results in Ref.\,\cite{Saito:2019rtg}, and show our result in secton\,\ref{100TeV}. Earlier projections from disappearing track searches from a compressed dark sector at 100\,TeV were carried out in\,\cite{Mahbubani:2017gjh}.

Our reconstruction proceeds as follows. At the generator level, $\slashed{p}_T$ is reconstructed as the vector sum of the $p_{T}$ of neutrinos, neutralinos and charginos since the tracklet $p_{T}$ is not used in the experimental reconstruction of missing transverse momenta. We reconstruct jets with {\tt FastJet3.1.3}\,\cite{Cacciari:2011ma} with $R=0.4$, and take as 
input all particles but muons, neutrinos, neutralinos and charginos with $c\tau>10$ mm. 

We use the benchmark SLHA files provided by the ATLAS collaboration and consider electroweak production 
of charginos via $pp\to\tilde{\chi}^{\pm}_{1}\tilde{\chi}^{0}_{1} j $ and 
$pp\to\tilde{\chi}^{+}_{1}\tilde{\chi}^{-}_{1} j$ at 13\,TeV in {\tt{MadGraph}}. We store the chargino 
decay vertex by setting the {\texttt{time\_of\_flight}} variable in the run card, decay the chargino in {\tt{Pythia}} and match our events with up to two extra partons using the MLM prescription\,\cite{Mangano:2006rw}. 

The following analysis selection criteria are imposed:

\begin{itemize}
\item Trigger : $\slashed{p}_T$ $> 140$\,GeV
\item Lepton veto : no electrons or muons
\item Jet $p_{T}$/$\Delta\phi$ : at least one jet with $p_{T} > 140$\,GeV, and $\Delta\phi$ between 
the $\slashed{p}_T$ vector and each of the up to four hardest jets with $p_{T}>50$\,GeV to be bigger than 1.0
\end{itemize}

In what follows, we use ``overall event level efficiency" to refer to the efficiency after these selection cuts. On top of these event selection requirements, we correct for detector effects and resolutions by multiplying the overall event level efficiency with the event efficiency provided by ATLAS in Table 2 of\,\cite{Aaboud:2017mpt}.\footnote{Note that Table 2 is provided and meant to be used for reinterpretation purposes, so we consider the event efficiencies and tracklet probability or $T_{P}$ in our validation, and later for our signal.}

\begin{figure}[t]
 \centering
 \includegraphics[width=0.6\textwidth]{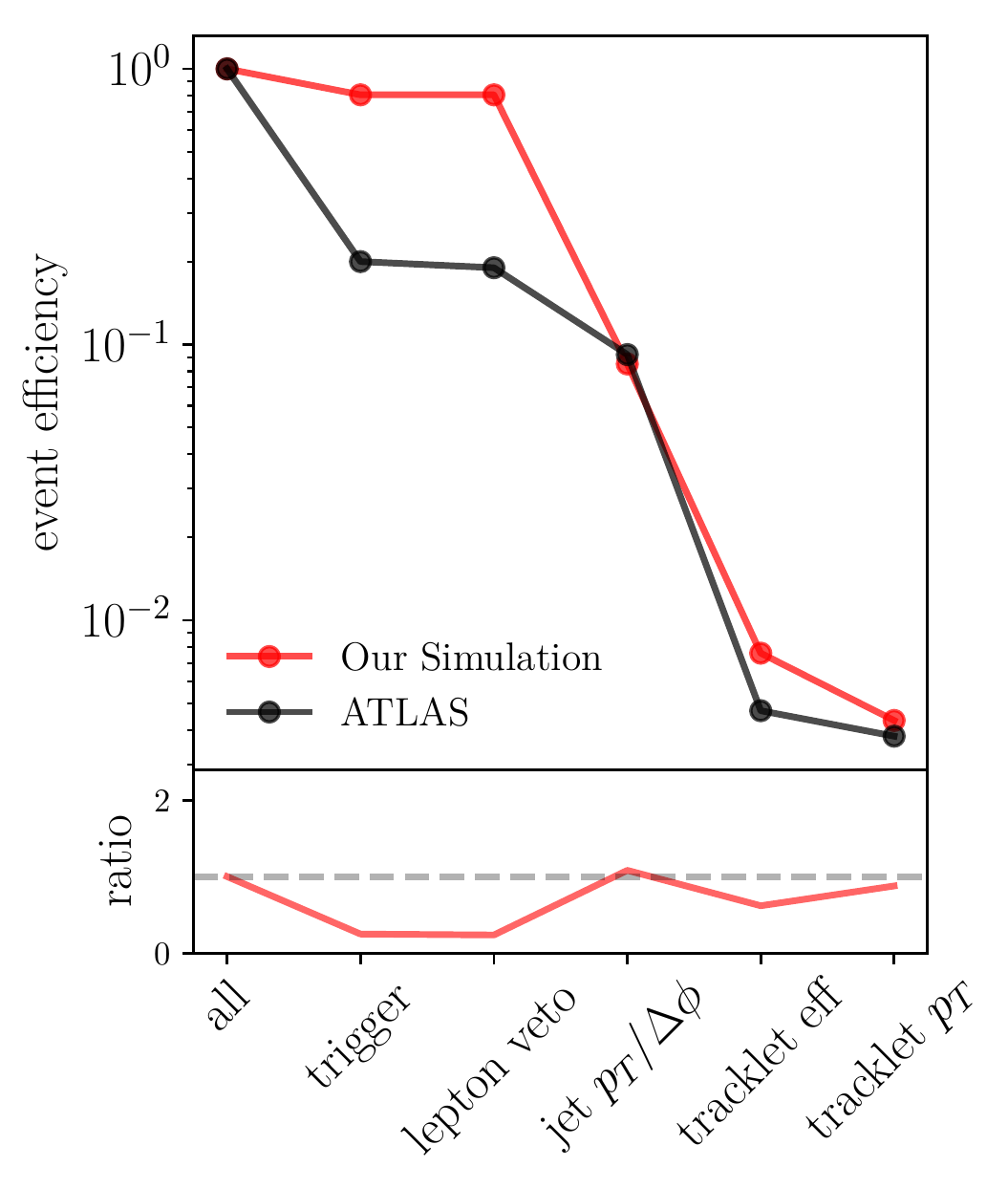}
 \caption{Validation of the ATLAS disappearing track search efficiency for a chargino 
 produced electroweakly with $(m_{\tilde{\chi}^{\pm}_{1}},c\tau_{\tilde{\chi}^{\pm}_{1}}) =$ ($400$\,GeV, $59.96$ mm). 
 The black curve corresponds to the ATLAS efficiency in Table 1 of Ref.\,\cite{Aaboud:2017mpt} and the red 
 curve corresponds to our simulation. The bottom rectangle shows the ratio of ATLAS's result to our estimate.}
 \label{fig:cutflow}
 \end{figure}

Then we proceed to select tracklets and require the following:

\begin{itemize}
\item Tracklet selection : at least one tracklet (generator-level chargino) with : 
\begin{itemize}
\item $p_{T}>20 $\,GeV and $0.1<|\eta|<1.9$
\item $122.5$ mm $<$ decay position $<295$ mm
\item $\Delta R$ distance between the tracklet and each of the up to four highest$-p_{T}$ jets with $p_{T}>50$\,GeV to be bigger than 0.4
\item  we apply the tracklet acceptance $\times$ efficiency map provided by ATLAS,\footnote{ We use auxiliary figure 9 of\,\cite{Aaboud:2017mpt} directly to account for the tracklet efficiency.} which is based on the decay position and $\eta$. This is applied to selected tracklets passing the above selections.
\end{itemize}
\item Tracklet $p_{T}$ : Select tracklets with $p_{T}>$ 100\,GeV.
\end{itemize}

In what follows, we use ``overall tracklet efficiency" to refer to the efficiency after these tracklet selection cuts. We correct our overall tracklet efficiency by a factor of 0.57, as presented in the last column of Table 2 from Ref.\,\cite{Aaboud:2017mpt}, which takes into account the experimental efficiency for reconstructing a tracklet with $p_T>100$\,GeV. 

In figure\,\ref{fig:cutflow}, we show the ATLAS result and our result by following the cutflow of Table 1 in Ref.\,\cite{Aaboud:2017mpt}. As can be seen from the plot, we reproduce the overall efficiency after all selection requirements are imposed. For $(m_{\tilde{\chi}^{\pm}_{1}},c\tau_{\tilde{\chi}^{\pm}_{1}})=(400\,\rm\,GeV, 59.96\,mm)$, the final efficiency for ATLAS is 0.38\% and we obtain 0.43\%.

\begin{figure}[t]
\centering{
  \begin{adjustbox}{max width = \textwidth}
\begin{tabular}{cc}
\includegraphics[scale=1.0]{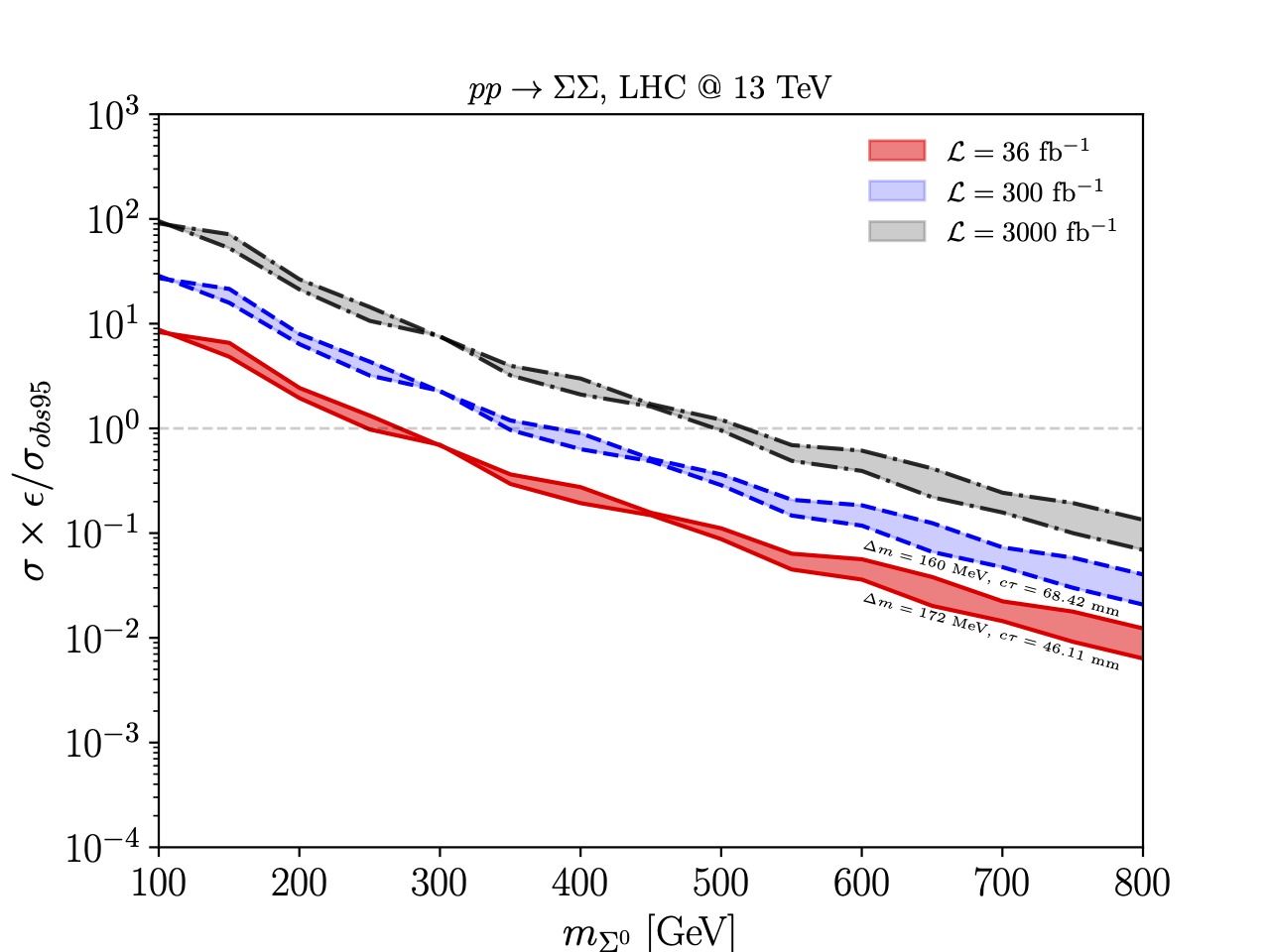}

\end{tabular}
  \end{adjustbox}}
\caption{{\color{black} $95\%$ CL exclusion limits at $\sqrt{s}=13$\,TeV LHC versus $m_{\Sigma^{0}}$. The red, the blue, and the gray bands correspond to $\mathcal{L}=36,\,300,\,$ and 3000\,$\rm fb^{-1}$ respectively. $\sigma_\mathrm{obs95}$ is the model-independent pair production cross section reported in Ref.\,\cite{Aaboud:2017mpt} for the chargino and $\sigma$ is the pair production cross section of the real triplet. The ``band'' feature for the latter results from the lifetime variation of $\Sigma^\pm$ at two loop. The lower boundary of each band corresponds to ($\Delta m = 172$ MeV, $c\tau = 46.11$ mm), while the upper boundary corresponds to ($\Delta m=160$\,MeV, $c\tau= 68.42$ mm).}}
\label{fig:13TeVlimit}
\end{figure}

\subsection{Sensitivity of the \ssm\, at the LHC}
\label{LHC13TeV}

For the \ssm, we apply the same selection cuts as discussed in section\,\ref{atlasval}, but we now replace the chargino with the charged $\Sigma$. ATLAS presents a model-independent observed limit at 95\% confidence level (CL) in table~4 of Ref.\,\cite{Aaboud:2017mpt}, $\sigma_\mathrm{obs95} = 0.22$ fb for $\mathcal{L}=36.1$ fb$^{-1}$ and $\sqrt{s}=13\rm\,TeV$.  We calculate our theoretical cross section $\sigma_\mathrm{theory}\equiv\sigma\times\epsilon$ for each mass point and compare that with $\sigma_\mathrm{obs95}$. If the ratio $\sigma_\mathrm{theory}/\sigma_\mathrm{obs95}>1$, then we consider the point to be excluded. {\color{black}The result is presented in figure\,\ref{fig:13TeVlimit}, where $\sigma$ is the pair production cross section of the real triplet particles. Note that, different from figure\,\ref{fig:sigma}, the cross section now has a ``band'' feature, which is a direct result of the lifetime variation of $\Sigma^\pm$ at two loops as discussed in the introduction. To be more specific, the lower boundary of each band in figure\,\ref{fig:sigma} corresponds to our benchmark point ($\Delta m = 172$ MeV, $c\tau = 46.11$ mm), while the upper boundary corresponds to ($\Delta m=160$\,MeV, $c\tau= 68.42$ mm).

Comparing the two benchmark points, one notices that, as the mass splitting decreases from 172 MeV to 160 MeV, the reach of the ATLAS DCT search increases. This is due to the fact that a smaller mass splitting leads to a longer lifetime for $\Sigma^\pm$ as can be seen from eq.\,\eqref{sig:decayrate}, and thus increases the sensitivity of the (HL-)LHC. To be precise, we find the efficiency of the recasted ATLAS analysis is indeed higher for the ($\Delta m=160$\,MeV, $c\tau= 68.42$ mm) benchmark than the other, raising to $0.4\%$ from $0.33\%$ for $\mathcal{L}=36\rm\,fb^{-1}$, i.e., the red band in figure\,\ref{fig:13TeVlimit}. Note also that currently the LHC with $\mathcal{L}=\rm36\,fb^{-1}$ already excludes a real triplet lighter than $\sim275$\,GeV for ($\Delta m=160$\,MeV, $c\tau= 68.42$\,mm), and lighter than $\sim248$\,GeV for ($\Delta m = 172$\,MeV, $c\tau = 46.11$\,mm).

For higher luminosities, in the optimistic case, one can obtain the corresponding exclusion limits by assuming that both $S$ and $B \propto {\mathcal L}$. As a consequence, the sensitivity ${S}/{\sqrt{B}}$ scales as $\sqrt{\mathcal L}$. In addition, one can also assume that $\sigma_\mathrm{obs95} \propto 1/\sqrt{\mathcal L}$ for the estimation. Based on these assumptions, and without considering any systematic uncertanties on the background, the (HL-)LHC would be able to explore the real triplet DM mass up to $m_\Sigma\sim 590$\,GeV ($\sim 745$\,GeV) and $\sim 535$\,GeV  ($\sim 666$\,GeV) for ($\Delta m=160$\,MeV, $c\tau= 68.42$ mm) and ($\Delta m=172$\,MeV, $c\tau= 46.11$ mm) respectively with $\mathcal{L}=300$\,fb$^{-1}$ ($3000$fb$^{-1}$).

The conclusion drawn from the optimistic approximation described above will change with the inclusion of background uncertainties. In general, performing a more careful extrapolations for LHC with higher luminosities is challenging for long-lived particle searches, particularly due to the difficulty in estimating instrumental backgrounds and uncertainties outside the experimental collaborations. A more conservative extrapolation procedure for DCT searches was studied in Ref.\,\cite{Belanger:2018sti}. There the authors argued that though the HL-LHC would be a much busier environment where backgrounds might not necessarily scale with the luminosity, it was also likely that the trigger upgrades/strategies could be improved to compensate for the larger backgrounds and therefore provide a larger signal statistics. Thus, we include the effect from systematics, and obtain more conservative exclusion limits based on the following assumptions: (1) The background uncertainty, dominated by systematics, remains constant when extrapolated to the LHC with higher luminosities (following the work in Ref.\,\cite{Belanger:2018sti}) and ; (2) The ATLAS DCT search reports a $30\%$ systematic uncertainty on the background at $36.1$ fb$^{-1}$, which will be used for our estimation on the sensitivity $S/{\Delta{B}}$ with $\Delta B=\sqrt{B+\delta B}$ and $\delta B=0.3B$; and (3) the number of signal and that of the background events scale with luminosity. 

In figure\,\ref{fig:13TeVlimit} we present our exclusions, with the inclusion of the $30\%$ systematic uncertainty for the LHC with high luminosities (blue and gray bands). The coverage of the LHC is now shifted to $\sim 382$\,GeV ($\sim 520$\,GeV) and $\sim 348$\,GeV  ($\sim 496$\,GeV) for ($\Delta m=160$\,MeV, $c\tau= 68.42$ mm) and ($\Delta m=172$\,MeV, $c\tau= 46.11$ mm) respectively at $\mathcal{L} = 300$\,fb$^{-1}$ ($3000$\,fb$^{-1}$). The weaker reach in this case is a direct result of the systematic uncertainty, and it changes the reach of the LHC with higher luminosities by a few hundred GeV compared with the optimistic approach.}

%%%%%%%%%
\subsection{Sensitivity of the \ssm\, at a $100$\,TeV $pp$ collider}\label{100TeV}
%%%%%%%%%
To assess the prospective sensitivity of a future 100\,TeV $pp$ collider, we rescale the leading jet $p_T$ and the $\slashed{p}_T$ cuts as suggested in\,\cite{Saito:2019rtg} with the following selections:

\begin{itemize}
\item Trigger : $\slashed{p}_T$ $> 1$\,TeV or $\slashed{p}_T$ $> 4$\,TeV depending on the benchmark as discussed below.
\item Lepton veto : no electrons or muons.
\item Jet $p_{T}$/$\Delta\phi$ : at least one jet with $p_{T} > 1$\,TeV, and $\Delta\phi$ between
the $\slashed{p}_T$ vector and each of the up to four hardest jets with $p_{T}>50$\,GeV to be bigger than 1.0.
\end{itemize}

The tracklet selection and tracklet $p_{T}$ cut remain the same as in the 13\,TeV case. The number of expected signal events at a 100\,TeV $pp$ collider with 30\,ab$^{-1}$ of luminosity are given in table\,\ref{table:Nsignal}, for two benchmarks. For the 1 (3)\,TeV benchmark point, the trigger threshold used is 1 (4)\,TeV. 

\begin{table}[t]
 \centering
 \begin{tabular}{c|c|c|c|c|c}
 Benchmark & $\sigma$ [pb] & $\epsilon$ & $S$ & $B$&  $S/\sqrt{B}$\\
   \hline
   \hline 
 $m_{{\Sigma}^{\pm}}=1.1$\,TeV,  $\overline{\mu} = 200$ &      $5.8 \times 10^{-2} $ & $3.17\times 10^{-4}$ & 553 & 673 &  21.3 \\
 $m_{{\Sigma}^{\pm}}= 1.1$\,TeV,  $\overline{\mu} = 500$  &      $5.8\times 10^{-2}$  & $3.17\times 10^{-4}$ & 553 & 8214 & 6 \\
\hline 
$m_{{\Sigma}^{\pm}}=3.1$\,TeV,  $\overline{\mu} = 200$  &       $9.4\times 10^{-4} $  & $ 4.69\times 10^{-4}$  & 13.3 & 1.9 & 9.6 \\
 $m_{{\Sigma}^{\pm}}=3.1$\,TeV,  $\overline{\mu} = 500$  &      $9.4\times 10^{-4}$   &  $4.69\times 10^{-4}$  & 13.3 & 27  & 2.6 \\
\hline
 \end{tabular}
 \caption{Cross section, overall event efficiency $ \epsilon$, number of expected signal (background) events $S$ ($B$) with $\mathcal{L}=30$ ab$^{-1}$ and significance $S/\sqrt{B}$ at a 100\,TeV $pp$ collider for two benchmarks with $(m_{{\Sigma}^{\pm}},c\tau_{{\chi}^{\pm}_{1}}) =$ ($1.1$\,TeV, $59.96$ mm) and 
 $(m_{{\Sigma}^{\pm}},c\tau_{{\Sigma}^{\pm}}) =$ ($3.1$\,TeV, $59.96$ mm), wherein the table $\bar{\mu}$ represents the average number of $pp$ interactions per bunch crossing. See the text for details. }
\label{table:Nsignal}
\end{table}

The authors in\,\cite{Saito:2019rtg} carefully considered the effect that multiple $pp$ collisions occurring simultaneously with a signal event (pileup) would have on the background. For each benchmark case, we adopt their fake tracklet background numbers considering the two different pileup scenarios  described in\,\cite{Saito:2019rtg}. We consider two values for $\overline{\mu}$ -- the average number of $pp$ interactions per bunch crossing -- the authors studied: $\overline{\mu}=200$ and $\overline{\mu}=500$. Values of $B$ in our Table \ref{table:Nsignal} are taken directly from Table 3 and 4 of\,\cite{Saito:2019rtg}. We then estimate the significance as $S/\sqrt{B}$. We conclude that a 100\,TeV $pp$ collider could discover $m_{\Sigma^0}=1\rm\,TeV$ and $m_{\Sigma^\pm}=1.1\rm\,TeV$ (significance larger than $5\sigma$) for both pileup scenarios. While with controlled pileup scenario ($\overline{\mu}<500$), the 100\,TeV collider could discover real triplet scalars with masses up to $m_{\Sigma^0}=3\rm\,TeV$ and $m_{\Sigma^\pm}=3.1\rm\,TeV$. As we discuss below, this reach would cover the entire DM viability range for portal coupling having a magnitude $\sim \mathcal{O}(1)$ and below.  We also stress that by optimizing the inner-tracker layout as done in Ref.\,\cite{Saito:2019rtg}, more optimal reach for the \ssm\, could be attained.

%%%%%%%%%%%%%%%%%%%%%%
\section{Triplet dark matter and direct detection}
\label{tripletdm}
%%%%%%%%%%%%%%%%%%%%%%

In secton\,\ref{collider}, we discussed \ssm\, DM searches from the DCT signature at the LHC and a 100\,TeV $pp$ collider. We find that, as shown in figure\,\ref{fig:13TeVlimit}, the 13\,TeV LHC can only reach the $m_\Sigma\sim\mathcal{O}(\rm100\,GeV)$ parameter space. However, as has been previously studied in Refs.\,\cite{Cirelli:2005uq,Cirelli:2007xd} in the case when $a_2=0$, $m_{\Sigma^0}$ has to be about $2.5\rm\,TeV$ in order to account for the entire DM relic density. Such a \,TeV scale is beyond the reach of the LHC, while for a 100 TeV $pp$ collider, the reach may extend to $m_{\Sigma^0}\sim 3$ TeV if pileup is under sufficient control.  Recall that for both colliders, the impact of the Higgs portal interaction with coefficient $a_2$ is minimal, except for the very light mass regime that is already excluded by LEP bounds. It is interesting, however, to study the interplay between the collider reach and DM dynamics in the presence of a non-vanishing Higgs portal interaction. As already observed in Ref.~\cite{Chao:2018xwz} for higher dimensional electroweak multiplet DM, the impact of the portal coupling on DM dynamics and direct detection sensitivity can be substantial. With this observation in mind, in this section, by taking non-zero $a_2$ into account, we discuss the parameter space where the \ssm\, can generate the measured DM relic density. We discuss constraints from DM direct detection at the end of this section.

%%%%%%%%%%%%%%%%%%%%%%
\subsection{Brief review of the Boltzmann equation with coannihilation}
%%%%%%%%%%%%%%%%%%%%%%
We assume that the DM particles stay in thermal equilibrium with the SM particles in the early universe. Due to the expansion of the universe, they eventually freeze out from the SM thermal bath when their annihilation rate is smaller than the Hubble rate. To understand how the DM abundance evolves with the expansion of the universe, one can solve the Boltzmann equation\footnote{For reviews on this topic, see, for example, Refs.\,\cite{Gondolo:1990dk,Kolb:1990vq}.}
\begin{equation}
\frac{dY}{dx}=\frac{1}{3H}\frac{ds}{dx}\langle\sigma v_{\rm M\o ller}\rangle_T(Y^2-Y^2_{\rm eq}),
\end{equation}
where $x\equiv m_{\rm DM}/T$, $Y_{\rm(eq)}\equiv n_{\rm(eq)}/s$, $s$ is the total entropy density of the universe, $n$ is the DM number density , $n_{\rm eq}$ is the number density  when the DM is in thermal equilibrium with the SM thermal bath, $T$ is the temperature of the SM thermal bath, $H$ is the Hubble rate, $v_{\rm M\o ller}\equiv\sqrt{(p_1.p_2)^2-m_1^2m_2^2}/(E_1 E_2)$ is the $\rm M\o ller$ velocity,\footnote{The subscripts 1 and 2 correspond to particle labels for the initial state of a general $2\to n$ scattering process.} and $\langle\sigma v_{\rm M\o ller}\rangle_T$ is the thermal-averaged annihilation cross section.

However, as discussed in secton\,\ref{setup},  $\Sigma^\pm$ is only 166\,MeV heavier than our DM candidate $\Sigma^0$. Consequently, coannihilation -- as first discussed in\,\cite{Griest:1990kh, Mizuta:1992qp} -- needs to be included.  To do so, we follow the general procedure described in Ref.\,\cite{Edsjo:1997bg} and rewrite the Boltzmann equation as
\begin{equation}
\frac{dY}{dx}=\frac{1}{3H}\frac{ds}{dx}\langle\sigma_{\rm eff} v_{\rm M\o ller}\rangle_T(Y^2-Y^2_{\rm eq}),\label{coanbzeq}
\end{equation}
where $ \langle\sigma_{\rm eff} v_{\rm M\o ller}\rangle_T$ can be written in a compact form:
\begin{equation}
\langle\sigma_{\rm eff} v_{\rm M\o ller}\rangle_T
\equiv
\frac{\displaystyle \int_{0}^{\infty} dp_{\rm eff} p_{\rm eff}^2 W_{\rm eff} K_1\left(\frac{\sqrt{\tilde{s}}}{T}\right)}
{\displaystyle m_{\rm DM}^4T\left[ \sum\limits_i \frac{g_i}{g_{\rm DM}} \frac{m_i^2}{m_{\rm DM}^2} K_2\left(\frac{m_i}{T}\right)\right]^2},
\end{equation}
and now with $Y\equiv{\sum\limits_i n_i}/{s}$ with $n_i$ the number density of species $i$ which is either the DM particle or other particles that will eventually decay into the DM particle, $K_{1(2)}$ the Bessel function of the first (second) kind, $\tilde{s}$ the Mandelstam variable, $g_i$ the number of degrees of freedom of species $i$, $m_{\rm DM}$ the mass of DM, $p_{\rm eff}=\sqrt{{\tilde{s}}/{4}-m_{\rm DM}^2}$ and $W_{\rm eff}=\sum\limits_{ij}({4p_{ij}^2}/{p_{\rm eff}})({g_ig_i}/{g_{\rm DM}^2})\sqrt{\tilde{s}}\sigma_{ij}$, where $p_{ij}={\sqrt{(\tilde{s}-(m_i+m_j)^2)(\tilde{s}-(m_i-m_j)^2)}}/({2\sqrt{\tilde{s}}})$ and the indices $i,\,j$ are the same as that in aforementioned $n_i$.

\begin{table}
\resizebox{\columnwidth}{!}{%
\begin{tabular}{l|llll}
\hline
Annihilation & \multicolumn{4}{c}{Coannihilation} \\
\hline
*$\Sigma^0\Sigma^0\to W^\pm W^\mp$ & $\Sigma^0\Sigma^\pm\to f\bar{f'}$ & *$\Sigma^\pm\Sigma^\mp\to f \bar{f}$ & *$\Sigma^\pm\Sigma^\mp\to h\gamma$ & $\Sigma^\pm\Sigma^\mp\to \nu\bar{\nu}$ \\ 
*$\Sigma^0\Sigma^0\to ZZ$ & $\Sigma^0\Sigma^\pm\to W^\pm Z$  & *$\Sigma^\pm\Sigma^\mp\to W^\pm W^\mp$ & *$\Sigma^\pm\Sigma^\mp\to hh$ & $\Sigma^\pm\Sigma^\pm\to W^\pm W^\pm$  \\ 
*$\Sigma^0\Sigma^0\to h h$ & $\Sigma^0\Sigma^\pm\to W^\pm\gamma$  & *$\Sigma^\pm\Sigma^\mp\to ZZ$ & $\Sigma^\pm\Sigma^\mp\to Z\gamma$ & \\ 
*$\Sigma^0\Sigma^0\to f\bar{f}$ & *$\Sigma^0\Sigma^\pm\to W^\pm h$  & *$\Sigma^\pm\Sigma^\mp\to Zh$ & $\Sigma^\pm\Sigma^\mp\to \gamma\gamma$ & \\ 
\hline
\end{tabular}%
}
\caption{Annihilation and coannihilation processes related to the DM relic density calculation, where $f=e,\mu,\tau,u,d,c,s,t,b$ and $\nu=\nu_e,\nu_\mu,\nu_\tau$. Processes starting with an asterisk (*) are $a_2$ dependent.}\label{tab:processes}
\end{table}

During the evolution of the universe, one can track the DM number density  by solving Eq.\,\eqref{coanbzeq}. To get the current DM relic density, noting that $Y\gg Y_{\rm eq}$ after the freeze-out, one has
\begin{equation}
\frac{1}{Y_0}=\frac{1}{Y_f}+\int_{x_0}^{x_f}dx\frac{1}{3H}\frac{ds}{dx}\langle\sigma_{\rm eff} v_{\rm M\o ller}\rangle_T\,,
\end{equation}
with $x_{f(0)}={m_{\rm DM}}/{T_{f(0)}}$, $T_{f(0)}$ the freeze-out (current) temperature, $Y_f$ the yield of DM at freeze-out and $Y_0$ the current yield of DM. Knowing $Y_0$, one can then compute the current dark matter relic density from
\begin{equation}
\Omega_{\rm DM}h^2=\frac{\rho_0}{\rho_c}h^2=\frac{8\pi Gs_0Y_0m_{\rm DM}}{3\times10^4}\,,
\end{equation}
with $\rho_c={3H^2}/({8\pi G})$ the critical density, $G$ the gravitational constant and $s_0$ the current entropy density.

%%%%%%%%%%%%%%%%%%%%%%
\subsection{Triplet dark matter relic density}
%%%%%%%%%%%%%%%%%%%%%%

In secton\,\ref{setup}, we have discussed that $\Sigma^0$ becomes our DM candidate in the presence of a discrete $Z_2$ symmetry and with $\langle\Sigma\rangle=0$. For the coannihilation processes discussed above, we list in table\,\ref{tab:processes} all relevant processes to be considered.

\begin{figure}[t]
\centering{
  \begin{adjustbox}{max width = \textwidth}
\begin{tabular}{cc}
\includegraphics[scale=0.35]{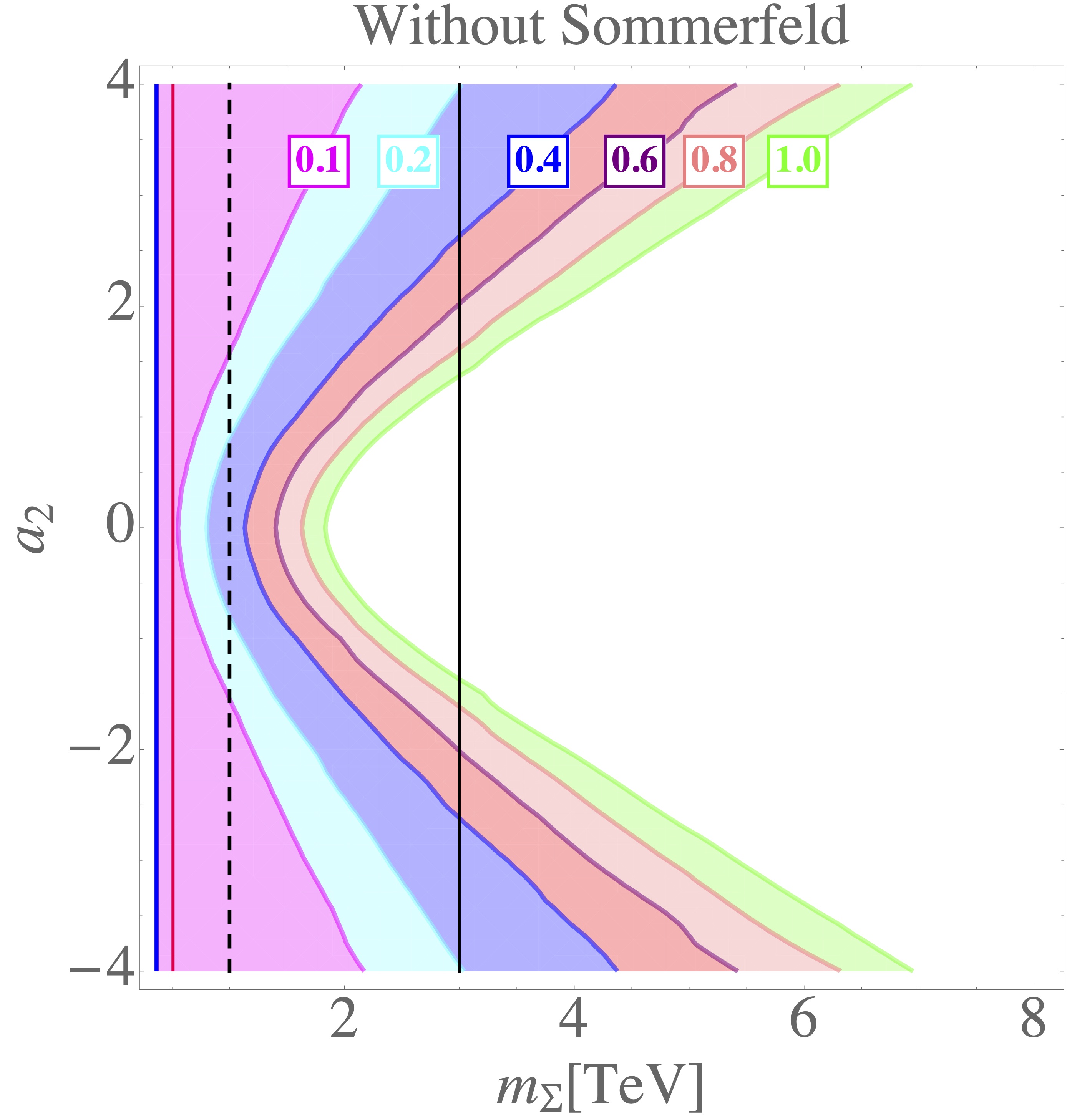} ~& ~ \includegraphics[scale=0.35]{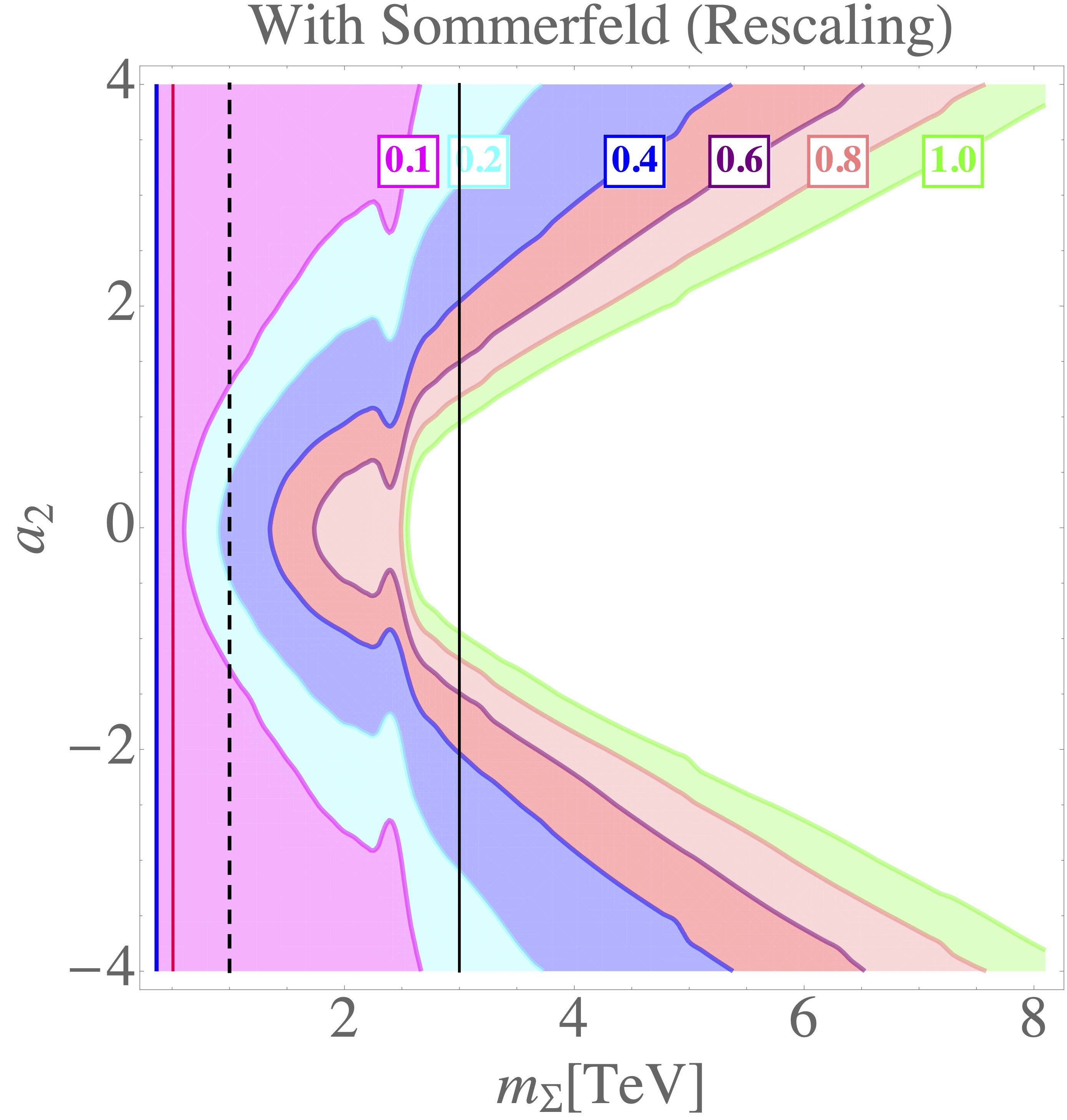}
\end{tabular}
  \end{adjustbox}}
\caption{Left panel: The parameter space that can explain current DM relic density without including the Sommerfeld effect. Numbers in boxes on the curves correspond to the fractions of \ssm\, contribution to the total DM relic density measured by Planck\,\cite{Aghanim:2018eyx}. The {\color{black}blue (red) vertical band} corresponds to the exclusion limit obtained {\color{black}in figure\,\ref{fig:13TeVlimit} with the inclusion of 30\% systematic uncertainties for higher luminosity LHC with $\mathcal{L}=300$\,fb$^{-1}$ and 3000\,fb$^{-1}$}, and the black dashed and solid lines correspond to the $\ge5\sigma$ discovery benchmark points we obtain in table\,\ref{table:Nsignal} for a future 100\,TeV $pp$ collider with $\mathcal{L}=30\rm\,ab^{-1}$. Right panel: Same as the left but with the Sommerfeld effect included.}\label{DMreach}
\end{figure}

%\footnotetext{{\color{black}Using the results with the inclusion of the 30\% systematic uncertainty will slightly shift these two bands to the left by a few hundred GeV.}}
In the analysis of Ref.~\cite{Cirelli:2007xd}, the authors obtained the \ssm\, relic density for $a_2=0$. We now turn our attention for the case with non-vanishing Higgs portal coupling. To study the effects of $a_2$ on the DM relic abundance, we first use {\tt LanHEP}\,\cite{Semenov:2014rea} to generate the model file. We then implement the model file in {\tt CalcHEP}\,\cite{Belyaev:2012qa} in order to calculate the annihilation and coannihilation cross sections.\footnote{We have checked that all the cross sections are in agreement with our hand-calculated results.} Then we use {\tt Mathematica} and {\tt Python} to solve the Boltzmann equation, Eq.\,\eqref{coanbzeq}.

Our results are shown in figure\,\ref{DMreach}, where we indicate the fraction of the relic density given by the $\Sigma^0$ (colored bands) in the ($m_{\Sigma^0}$, $a_2$) plane.\footnote{We have an agreement with Ref.\,\cite{Cirelli:2007xd} when $a_2=0$. For a general $a_2$, our relic density agrees with that calculated with MicrOMEGAs\,\cite{Belanger:2013oya}.} Numbers in boxes on the curves correspond to the fraction of the relic density comprised by the $\Sigma^0$, where the total relic density has been
measured by the Planck collaboration\,\cite{Aghanim:2018eyx}. {\color{black}The blue and the red vertical bands correspond to the collider exclusion limits we obtained from the DCT signature presented in figure\,\ref{fig:13TeVlimit}.} {Note that the exclusion limit from $\mathcal{L}=36\rm\,fb^{-1}$ in figure\,\ref{fig:13TeVlimit}, which is $\sim287\rm\,GeV$, is not explicitly shown in figure\,\ref{DMreach} as we assume $m_{\Sigma^0}>2m_{\rm SM}\,\rm$ so the real triplet can decay into all possible SM final states.} Black dashed and solid vertical lines in figure\,\ref{DMreach} are the discovery reach we obtain for a future 100\,TeV $pp$ collider in table\,\ref{table:Nsignal} under the optimistic pileup scenario.

From the left panel of figure\,\ref{DMreach}, one might na\"ively conclude that the LHC (HL-LHC) requires the triplet to contribute at least {\color{black}$\sim$10\%} of the total DM relic abundance from our study on the null DCT signature, as indicated by the {\color{black}blue and/or red vertical bands}.  One would further conclude that if the triplet is the only component of DM, $m_\Sigma$ is required to be $\gtrsim2$\,TeV ($m_\Sigma\simeq2\,$TeV when $a_2\simeq$0.), which is consistent with the previous studies\,\cite{Cirelli:2005uq,Cirelli:2007xd}. 

However, when $\Sigma^0$ is of $\mathcal{O}(\rm\,TeV)$, SM particles can be effectively taken as massless and non-perturbative contributions to the cross sections, also known as the Sommerfeld effect, need to be included\,\cite{Hisano:2006nn}. To do so, we first obtain the ratio of DM relic abundance between the two curves in the upper left panel of Figure 3 in Ref.\,\cite{Cirelli:2007xd}.  We then rescale our thermal-averaged cross sections in Eq.~\eqref{coanbzeq} by the corresponding factor for each $m_\Sigma$,\footnote{For $m_\Sigma\gtrsim 3\,$TeV, we make an extrapolation. And for each $m_\Sigma$, we use the same rescaling factor regardless of $a_2$.} and show our results in the right panel of figure\,\ref{DMreach}. The feature near $m_\Sigma\sim 2.5$ TeV indicates the existence of a DM bound state due to the attraction between DM particles from the Sommerfeld effect.\footnote{The dip always appears near $m_\Sigma\simeq2.5$\,TeV because we use the same factor obtained from Ref.\,\cite{Cirelli:2007xd} to rescale $\langle\sigma_{\rm eff}v_{\rm M\o ller}\rangle$.} Now due to the Sommerfeld enhancement of $\langle\sigma_{\rm eff}v_{\rm M\o ller}\rangle$, DM freezes out from the SM thermal bath at a later time and therefore results in a smaller DM relic density. Therefore, for a fixed $a_2$, the DM particle has to be heavier to freeze out earlier in order to explain the observed DM relic density. On the other hand, if the DM mass is fixed, then the coupling between DM and SM doublet has to decrease to have a smaller cross section for the DM to decouple from the SM thermal bath earlier. Note that now both the LHC and the HL-LHC  would require the triplet to contribute at least about 10\% of the total DM relic abundance if no disappearing track signature is observed at 95\% CL. For the \ssm\, to saturate the DM relic density, one must have $m_\Sigma\gtrsim2.5$\,TeV.

{\color{black}
%%%%%%%%%%%%%%%%%%%%%%
%\subsubsection{Comment on the Sommerfeld enhancement}
%%%%%%%%%%%%%%%%%%%%%%
\begin{figure}
\centering{
  \begin{adjustbox}{max width = \textwidth}
\begin{tabular}{cc}
\includegraphics[scale=0.3]{plots/WithSomRes} ~& ~ \includegraphics[scale=0.3]{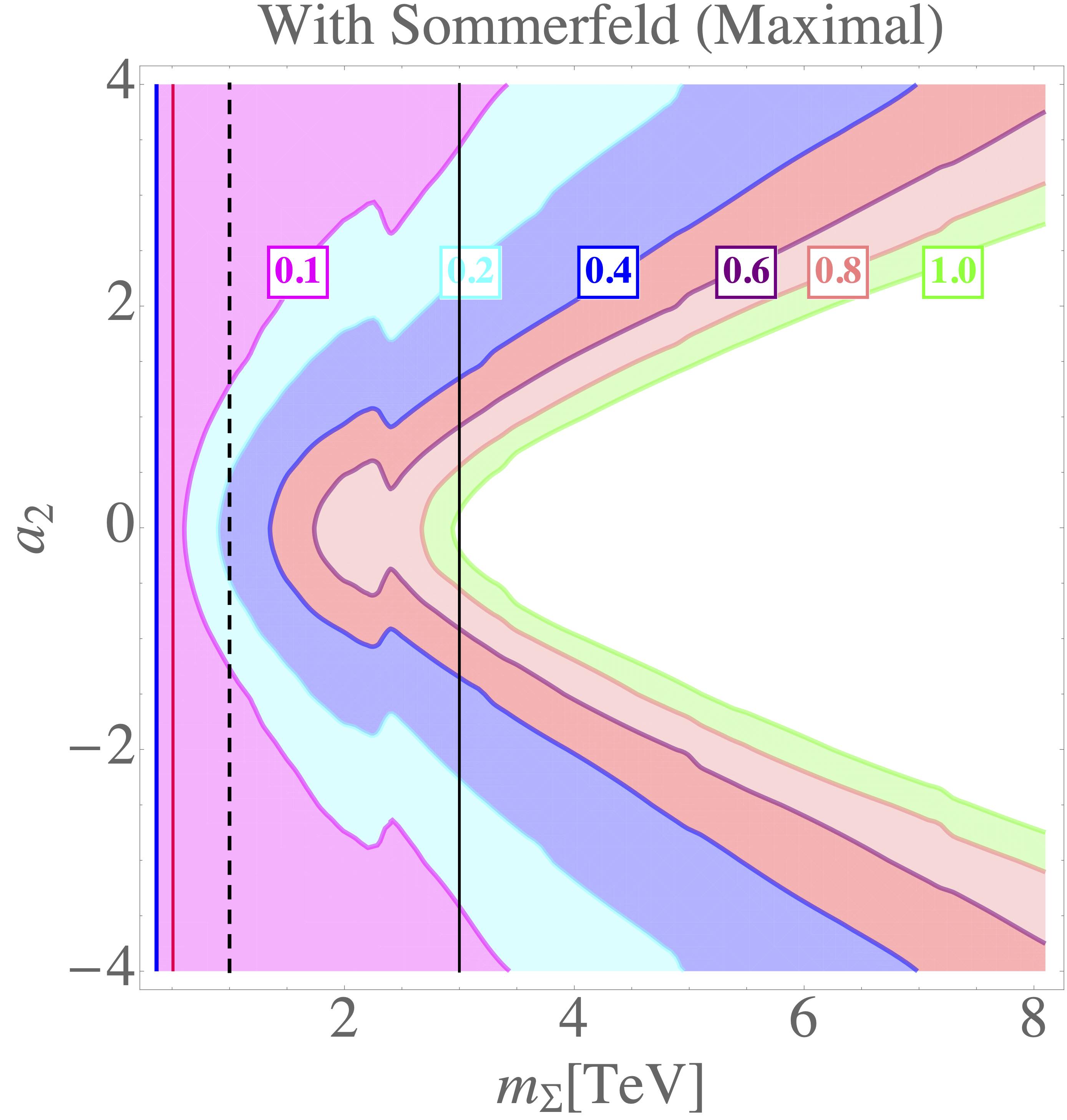}
\end{tabular}
  \end{adjustbox}}\caption{Left panel: The right plot of figure\,\ref{DMreach}, where the rescaling procedure is adopted for the Sommerfeld enhancement factor $S$ to obtain the DM relic density. Right Panel: To estimate the uncertainty from the rescaling, we assume $S$ stays at $S_{\rm max}$ when interpolation of the $S$ factor is not available. See the main text for a detailed discussion.}\label{NEWPLOT}
\end{figure}

As discussed above, the Sommerfeld enhancement corresponding to the right plot of figure\,\ref{DMreach} is obtained from a rescaling using the data from Ref.\,\cite{Cirelli:2007xd}. To be more specific, the rescaling is done through the following replacement
\begin{align}
\langle\sigma_{\rm eff} v_{\rm M\o ller}\rangle\to\langle S\sigma_{\rm eff} v_{\rm M\o ller}\rangle\label{SReplacement}
\end{align}
in eq.\,\eqref{coanbzeq}, where $S$ is the Sommerfeld factor interpolated from Ref.\,\cite{Cirelli:2007xd} when $m_\Sigma\lesssim3\rm\,TeV$ and extrapolated otherwise. Before discussing the uncertainties from the extrapolation, one general observation from eq.\,\eqref{SReplacement} is that, a larger $S$ factor would result in a larger DM annihilation cross section and thus a later freeze-out of DM and, thus, a smaller relic density for a given $m_\Sigma$.Therefore, a larger $m_\Sigma$ would be required to account for the observed DM relic density.

Now to estimate the uncertainty from the extrapolation for $m_\Sigma\gtrsim3\rm\,TeV$, we first examine the behavior of the $S$ factor for $500{\rm\,GeV}\le m_\Sigma\le8\rm\,TeV$, which is the mass range we consider in this work, using the Hulthen potential approximation. We find that there is indeed only one bound state near $m_\Sigma=$2\,TeV. Furthermore, the $S$ factor reaches its maximal value $S_{\rm max}$ near $m_\Sigma=2\rm\,TeV$ and then decreases dramatically for $m_\Sigma\ge2$\,TeV, meaning that the the Sommerfeld effect diminishes dramatically as well for $m_\Sigma\gtrsim2$\,TeV. In light of this, if one assumes that the $S$ factor stays at its maximal value $S_{\rm max}$ for $m_\Sigma\gtrsim2$\,TeV, then the required triplet mass to account for the correct DM relic density beyond 2\,TeV would also be the largest. The uncertainty from our rescaling procedure can then be obtained by comparing this largest value of $m_\Sigma$ with that shown in figure\,\ref{DMreach}.

%In light of this, the maximal uncertainty from the above-mentioned extrapolation can be estimated by assuming that the $S$ factor stays at its maximal value $S_{\rm max}$ for $m_\Sigma\gtrsim2$\,TeV.

Assuming $S=S_{\rm max}$ for $m_\Sigma\gtrsim2$\,TeV as discussed in the last paragraph and solving the Boltzmann equation again, we show the result in the right panel of figure\,\ref{NEWPLOT}, and for comparison, the result obtained from rescaling\footnote{{\color{black}Here, rescaling means the $S$ factor is obtained by interpolation from Ref.\,\cite{Cirelli:2007xd} for $m_\Sigma\lesssim3\rm\,TeV$ and extrapolation for $m_\Sigma\gtrsim3\rm\,TeV$.}} is shown in parallel in the left panel of figure\,\ref{NEWPLOT}. Now from the right panel of figure\,\ref{NEWPLOT}, one can find that to saturate the DM relic density from the real triplet, $m_\Sigma\gtrsim3$\,TeV is required, which is at the edge of the reachability of a future 100\,TeV collider and is about 500\,GeV larger than our conclusion drawn from the rescaling. However, as discussed above, since the $S$ factor actually decreases dramatically for $m_\Sigma\gtrsim2$\,TeV, this 500\,GeV shift shall be taken as the maximal correction. Another interesting point from the right plot of figure\,\ref{NEWPLOT} is that the DM relic density is now also less sensitive to $a_2$ as can be seen from the slopes of each colorful boundary.}

{\color{black}Another factor that could affect the DM relic density plots we obtain above comes from the bound state effect, which has been studied only recently in limited scenarios\,\cite{vonHarling:2014kha,An:2016gad,Cirelli:2016rnw,Mitridate:2017izz}. Contradictory conclusions on the bound state effects have been drawn for a U(1) model in Ref.\,\cite{vonHarling:2014kha,An:2016gad,Cirelli:2016rnw}, where Ref.\,\cite{An:2016gad} concluded that bound state effects were not important during thermal freeze-out of DM while Ref.\,\cite{vonHarling:2014kha,Cirelli:2016rnw} claimed the opposite. Bound state effects for WIMP DM were studied in Ref.\,\cite{MarchRussell:2008tu,Asadi:2016ybp,Shepherd:2009sa} by including only late time annihilations and leaving out the freeze-out processes. An effective field theory (EFT) approach to the wino scenario was developed in Ref.\,\cite{Braaten:2017gpq,Braaten:2017kci,Braaten:2017dwq}, and the simplest setup with a dark sector charged under $\rm SU(2)_L$ was recently considered in Ref.\,\cite{Smirnov:2019ngs}. Using the framework developed in Ref.\,\cite{vonHarling:2014kha,Cirelli:2016rnw}, the authors of Ref.\,\cite{Smirnov:2019ngs} also claimed the importance of the inclusion of the bound state effects especially near the unitarity bound, which, however, is not quite relevant to our study here due to the mass range we consider.

On the other hand, since the real triplet dark matter is very heavy, DM annihilation into energetic SM particles happening in the dark matter halo nearby could result in signals that can be detected by DM indirect detection though large uncertainties exist. The constraints from DM indirect detection could be strong and, thus, rule out part of our model parameter space. Actually, in the minimal dark matter (MDM) scenario where the only free parameter of the real triplet is $m_\Sigma$, DM relic density would require $m_\Sigma\simeq2.5$\,TeV, which has already been ruled out from DM indirect detection\,\cite{Cirelli:2005uq,Cirelli:2009uv,Cirelli:2015bda}. Interpreting this result in the real triplet model, it corresponds to an exclusion of the $[a_2=0,\,m_\Sigma\simeq2.5{\rm\,TeV}]$ point in figures\,\ref{DMreach} and \ref{NEWPLOT}. Due to the difference between the\,\ssm\,and the MDM models, the real triplet interacts differently with the SM particles compared with that in the MDM scenario. As a result of this difference, a plethora of the real triplet parameter space could still survive from DM indirect detection, as was also stated in Appendix B of Ref.\,\cite{Cirelli:2009uv}.

A more careful study on the Sommerfeld effect and a full study on the bound state effects and the constraints from DM indirect detection would be interesting and is highly non-trivial. We thus leave it for a future project.}

%%%%%%%%%%%%%%%%%%%%%%
\subsection{Triplet dark matter direct detection}
%%%%%%%%%%%%%%%%%%%%%%

The $\Sigma$ can interact with SM particles via the $a_2$ term in the Lagrangian. Again, as noted previously for higher dimensional electroweak multiplet DM, the presence of non-vanishing Higgs portal interaction can significantly enhance the cross section for DM-nucleus scattering\cite{Chao:2018xwz}. Therefore, we anticipate that for non-vanishing $a_2$,  the spin-independent (SI) cross section from dark matter scattering off nucleons can be severely constrained from deep underground experiments such as LUX\,\cite{Akerib:2016vxi}, PandaX-II\,\cite{Cui:2017nnn} and XENON1T\,\cite{Aprile:2018dbl}. 

Historically, the SI cross section was first studied by using the effective Lagrangian between DM and light quarks and gluons by Drees and Nojiri\,\cite{Drees:1993bu} and then followed by Refs.\,\cite{Jungman:1995df,Hisano:2010fy,Hisano:2011cs,Hisano:2014kua,Hill:2014yka,Hill:2014yxa,Chao:2018xwz}. Here we adopt the formula in Ref.\,\cite{Chao:2018xwz} for $m_\Sigma\gg m_W\gg m_q$ ($q=u,d,s$) and write the SI cross section as
\begin{align}
\begin{split}
&
\sigma_{\rm SI}=\frac{\mu^{2}}{4 \pi} \frac{m_{N}^{2}}{M_{\Sigma}^{2}}\left(f_{N} \frac{a_2}{m_{h}^{2}}+\frac{3}{4} f_{T} f_{N}^{\mathrm{PDF}}\right)^{2},\label{sixsection}\\
&\mbox{where}~
f_{T}=\frac{\alpha_{2}^{2}}{4 m_{W}^{2}} \left\{\omega \ln \omega+4+\frac{(4-\omega)(2+\omega) \arctan 2 b_{\omega} / \sqrt{\omega}}{b_{\omega} \sqrt{\omega}}\right\},
\end{split}
\end{align}
with $\mu=\frac{m_N m_\Sigma}{(m_N+m_\Sigma)}$ the reduced mass, $\omega=\frac{m_W^2}{m_\Sigma^2}$, $m_N$ the nucleon mass, $f_{N}$ the SI effective coupling with $f_N\simeq0.287(0.084)$ for $N=p\,(n)$\,\cite{Belanger:2013oya},\footnote{The authors in Ref.\,\cite{Alarcon:2011zs,Alarcon:2012nr} obtained $f_N=0.28(7)$ from pion-nucleon scattering and pionic-atom data using chiral effective field theory, which is consistent with the number we use here.} $f_{N}^{\mathrm{PDF}}=0.526$\,\cite{Hisano:2015rsa} the second moment of nucleon parton distribution function (PDF) and $f_T$ the effective coupling of the twist-two operator in the effective Lagrangian.

\begin{figure}[t]
\centering{
  \begin{adjustbox}{max width = \textwidth}
\begin{tabular}{cc}
\includegraphics[scale=0.32]{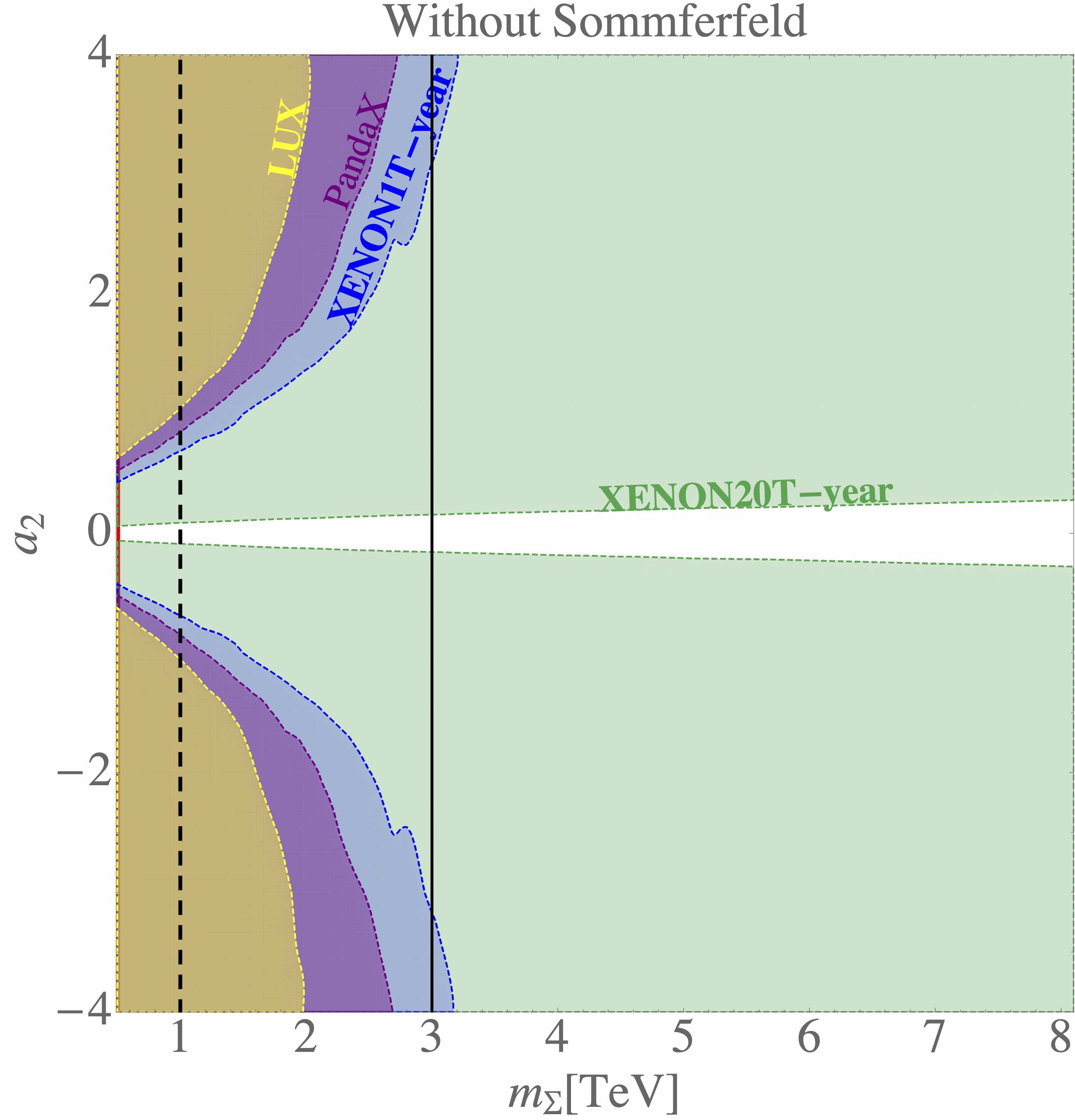} ~& ~ \includegraphics[scale=0.32]{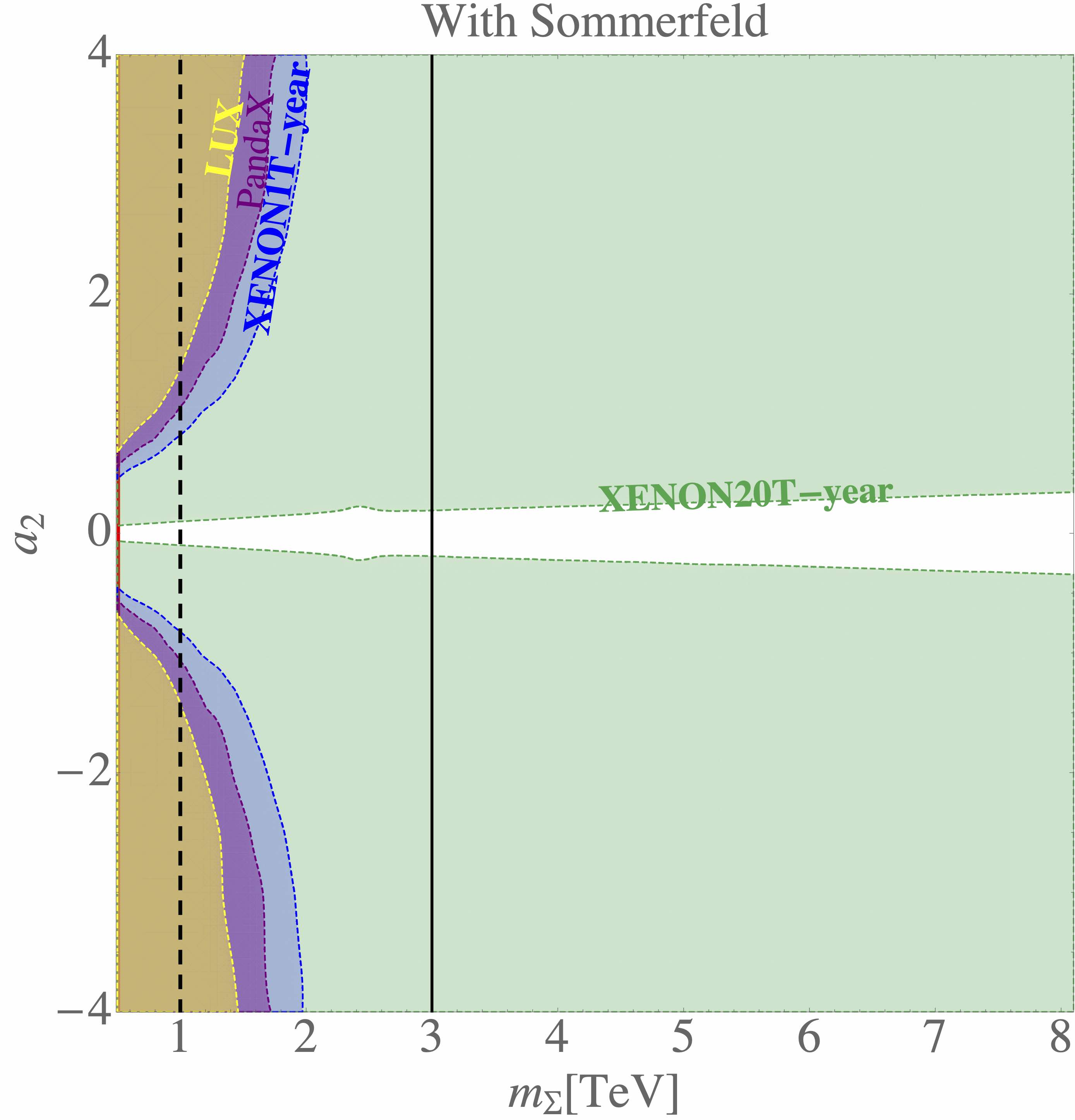}
\end{tabular}
  \end{adjustbox}}
\caption{Scaled spin-independent cross section $\sigma_{\rm SI}^{\rm scaled}$ on the $a_2$-$m_\Sigma$ plane. Left panel: Exclusion regions when the Sommerfeld effect is not included, where the yellow region is the constraint from LUX\,\cite{Akerib:2016vxi}, purple from PandaX-II\,\cite{Cui:2017nnn}, blue from XENON1T\,\cite{Aprile:2018dbl} and green from the projected XENON20T. The vertical lines have the same meaning as in figure\,\ref{DMreach}.\protect\footnotemark Right panel: Same as the left but with the Sommerfeld effect included.}\label{dmdirdet}
\end{figure}

\footnotetext{{\color{black}Note that constraints from the LHC with $\mathcal{L}=300\rm\,fb^{-1}$ is absent in this figure since the mass range shown here is [500\,GeV, 8\,TeV], which is beyond the reach of the LHC with $\mathcal{L}=300\rm\,fb^{-1}$.}}

Using Eq.\,\eqref{sixsection} and recasting constraints from LUX, PandaX-II and XENON1T onto the $a_2$-$m_\Sigma$ plane, we calculate the scaled SI cross section, which is defined as 
\begin{equation}
\sigma_{\rm SI}^{\rm scaled}\equiv \sigma_{\rm SI}\times\frac{\Omega h^2}{(\Omega h^2)_{\rm Planck}}\quad {\rm ~with~} (\Omega h^2)_{\rm Planck}=0.1198\text{\,\cite{Aghanim:2018eyx}},
\end{equation}
and plot constraints from those experiments in figure\,\ref{dmdirdet} {\color{black}based on the criterion that $\sigma_{\rm SI}^{\rm scaled}>\sigma_{\rm SI}^{\rm exp.}$ will be excluded}. In both plots, the yellow region corresponds to the exclusion limit from LUX, the purple region is excluded from PandaX-II, the blue region is excluded from XENON1T, and the green region is the projected exclusion limit from XENON20T\,\cite{Aprile:2018dbl}. Several points are worth stressing:
\begin{itemize}
\item Among the considered underground experiments, XENON1T gives the most stringent constraint in the $a_2$-$m_\Sigma$ parameter space. As can be seen from the left (right) panel of figure\,\ref{dmdirdet}, XENON1T excludes $m_\Sigma\lesssim3.2~(2)$\,TeV for $|a_2|\simeq4$ when the Sommerfeld effect is not (is) included. However, for $|a_2|\lesssim0.25$, the triplet can be as light as $\mathcal{O}(100\,\rm\,GeV)$, but cannot saturate the current DM relic density, as seen in figure\,\ref{DMreach}.

\item {From figure\,\ref{DMreach} and figure\,\ref{dmdirdet}, we conclude that, with or without including the Sommerfeld enhancement effect, current DM direct detection still permits the real triplet to be the sole DM candidate.} Moreover, looking into the future, XENON20T will cover almost the entire parameter space of the \ssm. Therefore, it is very promising for XENON20T to directly observe the signal of a real triplet DM. 

\item As can be seen from the right panel of figure\,\ref{dmdirdet}, when the Sommerfeld effect is included, exclusion regions from deep underground experiments all shrink. The reason is that, after including the Sommerfeld effect, the theoretical DM relic density $\Omega h^2$ at the same point in the plane becomes smaller due to a later freeze-out, as also seen in figure\,\ref{DMreach}. As a result, $\sigma_{\rm SI}^{\rm scaled}$ also becomes smaller and the corresponding parameter space is less constrained. 

\item In both plots of figure\,\ref{dmdirdet}, {\color{black}the blue (red) vertical band} corresponds to the exclusion limit we obtain in figure\,\ref{fig:13TeVlimit} for the LHC with $\mathcal{L}=300~(3000)\rm\,fb^{-1}$ and the black dashed and solid lines correspond to the $\ge5\sigma$ discovery benchmark points we have in table\,\ref{table:Nsignal} for a future 100\,TeV $pp$ collider with $\mathcal{L}=30\rm\,ab^{-1}$. As one may see, the LHC can only reach the low mass regime up to about 1\,TeV,  well below the  mass required to saturate the relic density.  However, a future 100\,TeV $pp$ collider will reach further into the\,TeV regime. In particular, in the white regions where XENON20T loses its sensitivity when $|a_2|\lesssim0.1$, future hadron colliders will be the key for model discovery.

\item Theoretical constraints on the triplet potential including bounded from below, unitarity and perturbativity have been studied in Ref.\,\cite{Khan:2016sxm,Chabab:2018ert} and recently reviewed in Ref.\,\cite{Bell:2020gug}. {For the parameter space we consider here, perturbativity and perturbative unitarity are satisfied with a cutoff scale $\Lambda\gtrsim10^6$\,GeV ($\Lambda\simeq10^6$\,GeV for $a_2\simeq4$) as implied in the right panel of Figure 1 in Ref.\,\cite{Bell:2020gug}. Requiring perturbativity and perturbative unitarity up to a higher scale will result in a smaller upper bound on $a_2$ than what we choose in figure\,\ref{dmdirdet}.}
\end{itemize}

%%%%%%%%%%%%%%%%%%%%%%
\section{Conclusions}\label{sec:summary}
%%%%%%%%%%%%%%%%%%%%%%

We consider a simple extension of the SM with a real triplet $\Sigma$, which transforms as (1,3,0) under the SM gauge group. The charged triplet component, $\Sigma^\pm$, has a degenerate mass as the neutral component, $\Sigma^0$, at tree level, but receives electroweak radiative corrections to become 166\,MeV heavier than $\Sigma^0$ at the one-loop level, {\color{black}and a further few MeV if two-loop corrections are also included}. The neutral component $\Sigma^0$ becomes stable and a dark matter candidate if $\langle\Sigma\rangle=0$ and an additional discrete $Z_2$ symmetry is imposed. Due to the small mass splitting between $\Sigma^\pm$ and $\Sigma^0$, $\Sigma^\pm$ becomes relatively long-lived, with the dominant decay channels being $\Sigma^\pm\to\pi^\pm\Sigma^0$. The pion in the final state is too soft to be reconstructed in colliders. Therefore, once $\Sigma^\pm$ is produced at colliders, a disappearing track, to which the LHC is currently sensitive, can be observed.

In this paper, the disappearing track signature at the LHC and a hypothetical 100\,TeV $pp$ collider is studied. We reproduce the ATLAS disappearing track efficiency in Ref.\,\cite{Aaboud:2017mpt}, as shown in figure\,\ref{fig:cutflow}, and then apply the same setup to our model. Our simulation result for the \ssm\, is shown in figure\,\ref{fig:13TeVlimit}. {\color{black}We find that, using the disappearing track signature, the 13\,TeV LHC excludes a real triplet lighter than 275 (248)\,GeV, 590 (535)\,GeV and 745 (666)\,GeV for $\mathcal{L}=\rm36\,fb^{-1},\,300\,fb^{-1},\,3000\,fb^{-1}$ and $\Delta m=160\,(172)$\,MeV, respectively. On the other hand, with the inclusion of a 30\% systematic uncertainty for the LHC with high luminosities, coverage of the LHC is shifted to 382 (348)\,GeV and 520 (496)\,GeV for $\mathcal{L}=\rm\,300\,fb^{-1},\,3000\,fb^{-1}$ and $\Delta m=160\,(172)$\,MeV, respectively. Due to the difficulty in estimating the instrumental backgrounds from simulation, our projection to the HL-LHC only serves as a rough estimation for the future HL-LHC.} We also extrapolate the disappearing track efficiency for a 100\,TeV $pp$ collider and study the reach at two benchmark points representative of FCC-pileup conditions, following the detailed background study by the authors in\,\cite{Saito:2019rtg}. We find that, even though the LHC can only cover the $\mathcal{O}(\rm 100\,GeV)$ regime of the \ssm, a 100\,TeV $pp$ collider will potentially be able to reach the TeV regime of the parameter space, provided that future pileup levels remain low, as shown in table\,\ref{table:Nsignal}. We stress that this is a motivation for more detailed experimental studies at 100\,TeV, as they can alter the potential of discovering the \ssm\, significantly.

On the other hand, understanding the particle nature of DM has been a profound problem in modern particle physics. It has been known that to explain the current DM relic density measured by the Planck satellite, the triplet needs to be heavier than about 2.5\,TeV, way above the scale that can be reached by the LHC. However, the triplet DM can interact with the SM particles via a Higgs portal coupling $a_2$ and the effects can be observed through DM direct detection from nucleon recoils. We investigate the constraints from LUX, PandaX-II, XENON1T and the projected XENON20T, and show our result in figure\,\ref{dmdirdet}. We find that currently XENON1T gives the most stringent constraint on our model parameter space and, for example, has excluded a triplet lighter than $\sim3$\,TeV for $|a_2|\simeq4$. In the future, XENON20T will be able to cover almost the entire parameter space of the \ssm\, model, except for $|a_2|\lesssim0.1$, where the interaction between the DM and the nucleons becomes too weak for deep underground detectors to have any sensitivity. Fortunately, a 100\,TeV $pp$ collider could have the chance to explore this region in the future.

\vspace{0.5cm}

{\it{Note added:}}  While finishing up this work, Ref.\,\cite{Bell:2020gug} appeared and also focused on the phenomenology of the \ssm. The main difference is that in our work, we focus on the $\Sigma^{0}$ dark matter scenario only (our neutral triplet scalar is fully stable). We compute the relic density including {\color{black}an approximate estimation from the Sommerfeld effect, while it was ignored in Ref.\,\cite{Bell:2020gug}}. For the collider analysis, we take a different approach. We implement all cuts and corrections for the disappearing track search and project to the HL-LHC -- {The current ATLAS limit we obtain is consistent with that in Ref.\,\cite{Bell:2020gug}. We also discuss discovery prospects at a future 100\,TeV $pp$ collider.} Our work complements that in Ref.\,\cite{Bell:2020gug}.

\acknowledgments{The authors thank R. Sawada, J. Zurita, F. Rojas, J.-H. Yu, H. Patel, A. Belyaev, O. Mattelaer and L. Friedrich for helpful discussions. YD also thanks the Institute of Theoretical Physics, Chinese Academy of Science --where a portion of this work was completed -- for hospitality and local support. CWC was supported in part by the Ministry of Science and Technology (MOST) of Taiwan under Grant Nos. MOST-104-2628-M-002-014-MY4 and MOST-108-2811-M-002-548. GC acknowledges support from grant No. MOST-107-2811-M-002-3120 and ANID/FONDECYT-Chile grant No. 3190051. YD, KF, and MJRM were supported in part under U.S. Department of Energy contract No. DE-SC0011095. MJRM was also supported in part under National Science Foundation of China grant No. 19Z103010239. KF was also supported by the LANL/LDRD Program.}

%%%%%%%%%%%%%%%%%%%%%%%%%
\bibliographystyle{JHEP}

\end{document}